\documentclass[pdftex]{pasj01}
\begin{document} 
\Received{}%{yyyy/mm/dd}
\Accepted{}%{yyyy/mm/dd}
%\Published{yyyy/mm/dd}

\title{The Thirty Millimeter Telescope}

%%% begin:list of authors
% Do NOT capitalize all letters in "textsc".
\author{Yoshifusa \textsc{Ita}\altaffilmark{1}%
%\thanks{Example: Present Address is xxxxxxxxxx}
}
\altaffiltext{1}{Astronomical Institute, Graduate School of Science, Tohoku University, 6-3 Aramaki Aoba, Aoba-ku, Sendai, Miyagi 980-8578, Japan}
\email{yita@astr.tohoku.ac.jp}

\author{Takashi \textsc{Ichikawa}\altaffilmark{1}}
%\altaffiltext{2}{B-Address of Institute}
%\email{bbbbb@xxx.xxx.xx.xx}

\author{Hironori \textsc{Tsutsui}\altaffilmark{2}}
\altaffiltext{2}{Subaru Telescope Okayama Branch Office, National Astronomical Observatory of Japan, 3037-5 Honjo, Kamogata, Asakuchi, Okayama, 719-0232, Japan}
\author{Takumi \textsc{Hanaue}\altaffilmark{1}}
\author{Takahiro \textsc{Komiyama}\altaffilmark{1}}
\author{Hiroki \textsc{Onozato}\altaffilmark{3,1}}
%\altaffiltext{3}{Nishi-Harima Astronomical Observatory, Center for Astronomy, Institute of Natural and Environmental Sciences, University of Hyogo, 407-2 Nishigaichi, Sayo-cho, Sayo-gun, Hyogo 679-5313, Japan}
\altaffiltext{3}{National Astronomical Observatory of Japan, 2-21-1 Osawa, Mitaka, Tokyo 181-8588, Japan}
\author{Atsushi \textsc{Iwamatsu}\altaffilmark{1}}
\author{Ryosuke \textsc{Morita}\altaffilmark{1}}
\author{Yuta \textsc{Habasaki}\altaffilmark{1}}
\author{Ryuto \textsc{Amemiya}\altaffilmark{1}}
\author{Miho \textsc{Hanawa}\altaffilmark{1}}
\author{Kenshi \textsc{Yanagisawa}\altaffilmark{3,2}}
\author{Hideyuki \textsc{Izumiura}\altaffilmark{2}}
\author{Yoshikazu \textsc{Nakada}\altaffilmark{4}}
\altaffiltext{4}{Institute of Astronomy, Graduate School of Science, The University of Tokyo, 2-21-1 Osawa, Mitaka, Tokyo, 181-0015, Japan}

%\email{ccccc@xxx.xxx.xx.xx}
%%% end:list of authors

%% `\KeyWords{}' always has to be placed before `\maketitle'.
 %Do NOT move this preamble from here!

\maketitle

\begin{abstract}
A near-infrared telescope with an effective aperture diameter of thirty millimeters has been developed. The primary objective of the development is to observe northern bright stars in the $J$, $H$, and $K_{\rm s}$ bands and provide accurate photometric data on those stars. The second objective is to repeatedly observe a belt-like region along the northern Galactic plane ($|b| \le 5^\circ$ and $\delta \ge -30^\circ$) to monitor bright variable stars there. The telescope has been in use since December 2016. The purpose of this paper is to describe the design and operational performances of the telescope, photometric calibration methods, and our scientific goals. We show that the telescope has the ability to provide photometry with an uncertainty of less than 5\% for stars brighter than 7, 6.5, and 6~mag in the $J$, $H$, and $K_{\rm s}$ bands, respectively. The repeatability of the photometric measurements for the same star is better than 1\% for bright stars. Our observations will provide accurate photometry on bright stars that are lacking in the Two Micron Sky Survey and the Two Micron All-Sky Survey. Repeated observations at a good cadence will also reveal their nature of the variability in the near-infrared.
\end{abstract}

%\linenumbers

\section{Introduction}
\begin{figure}
\begin{center}
\includegraphics[scale=0.43,angle=0]{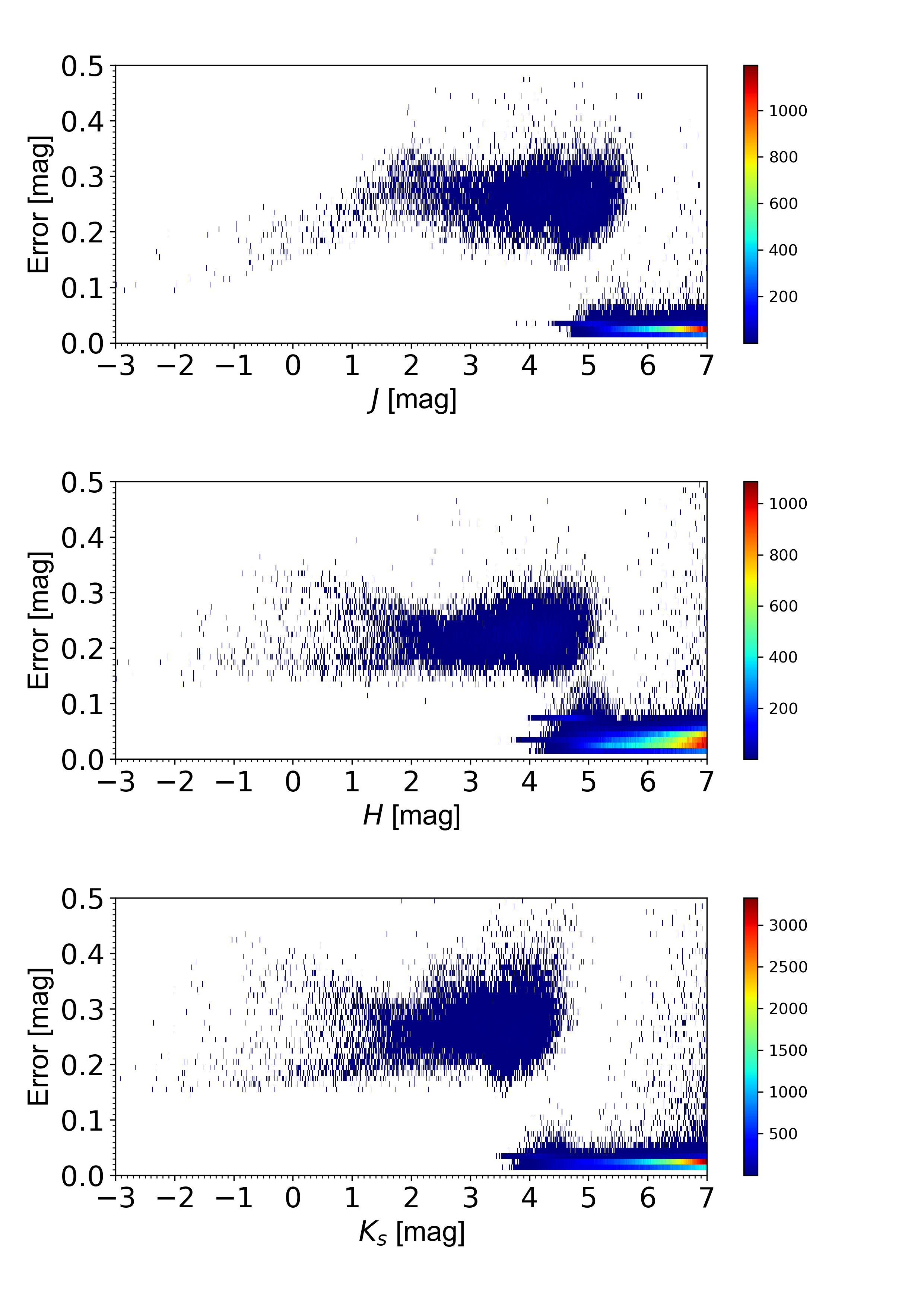} 
\end{center}
\caption{Heatmap of the 2MASS apparent magnitudes and their associated errors for $J$, $H$,  and $K_{\rm s}$ bands. Sources brighter than 7~mag are shown. The width of the bin is chosen to be 0.01~mag for both x and y axes. The color bar shows the number of sources in the 0.01 $\times$ 0.01~mag bin.}
\label{errorplot}
\end{figure}

\begin{figure}
\begin{center}
\includegraphics[scale=0.3,angle=0]{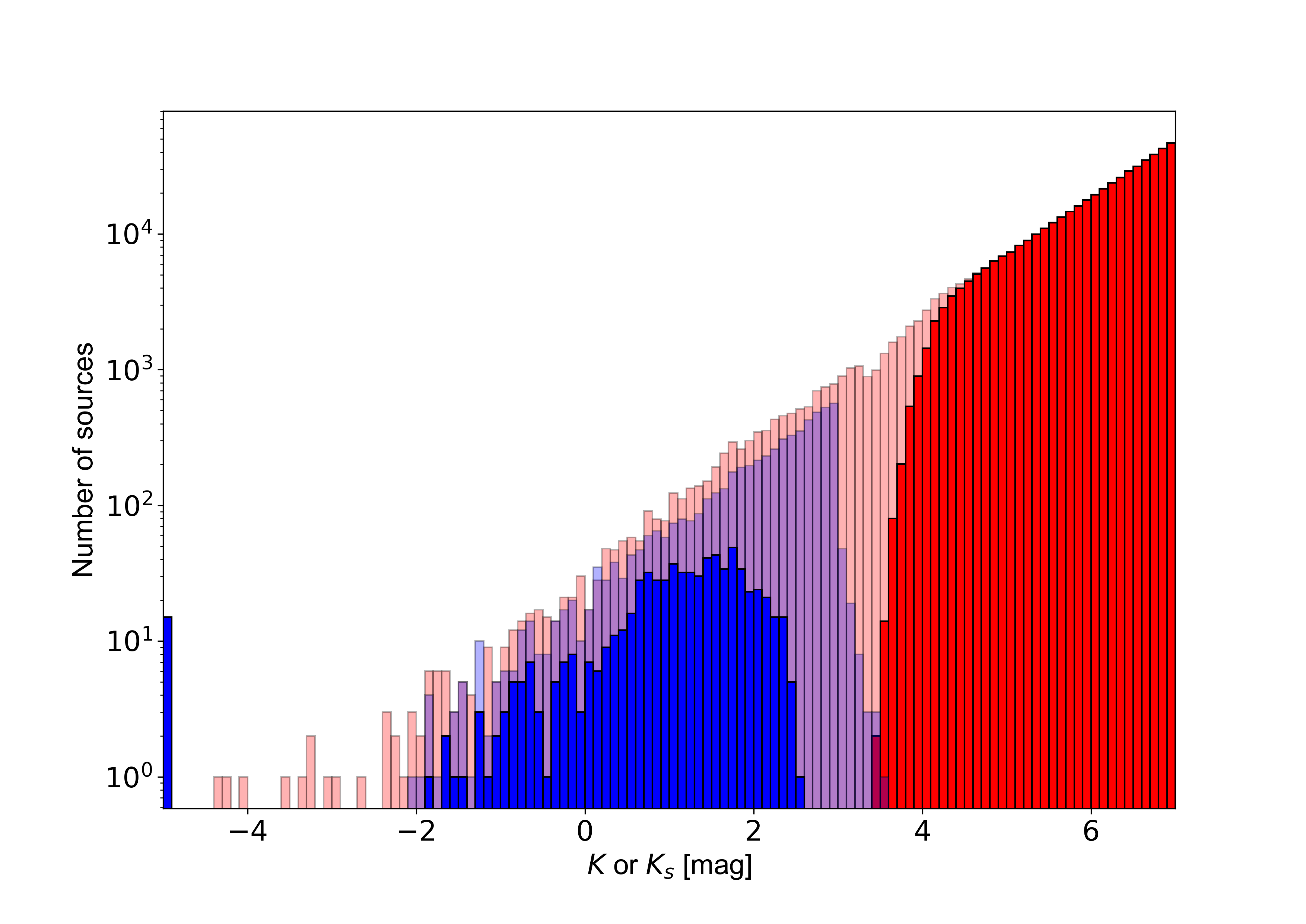} 
\end{center}
\caption{Histogram of the number of sources at $K_{\rm s}$ in the 2MASS (red), and at $K$ in the TMSS (blue) catalogs. Sources brighter than 7~mag are shown. Transparent colors show all sources in each catalog, and the opaque colors indicate sources with a photometric error smaller than 0.05~mag (corresponding to the signal to noise ratio of $\sim$20).}
\label{errorplot2}
\end{figure}

We human's insatiable curiosity reaches at a point to build gigantic telescopes that have an effective mirror diameter of over thirty meters (The Thirty Meter Telescope, TMT: \cite{sanders2013}, The Extremely Large Telescope, ELT: \cite{neichel2018}). The larger the aperture size of the telescope, the fainter celestial objects we can see, which helps to expand the horizons of human knowledge of the universe. On the other hand, telescopes with large apertures have difficulties in observing bright objects due to the saturation of detectors. Ironically enough, faint stars have better photometric accuracy than bright ones especially in near-infrared (NIR) at this time (see Figures~\ref{errorplot} and \ref{errorplot2}). However, is there any good reason to ignore or think lightly of bright stars? Bright stars are generally close to us, and therefore, their detailed follow-up observations are easy. There is a good chance that we may be missing something important that lies near at hand. 

Let us take a look at the current observational status of bright stars. Amateur astronomers have been playing an important role in providing observational data on bright stars in the optical. Their photometry data (both for single-epoch and multi-epoch observations) have been used for many scientific studies, especially for variable stars (e.g., \cite{molnair2019}). On the other hand, infrared data on stars brighter than $\sim$5.5 mag, which are too bright for modern infrared telescopes with sensitive detectors, are scarce even today. Technical aspects, such as requirement of cryogenic operation as well as the lack of information on driving infrared detectors and a know-how to operate them and to analyze the data output, hinder amateur astronomers from participating in NIR observations.

A new era of infrared astronomy was opened by the Two-Micron Sky Survey (TMSS) by \citet{neugebauer1969}. They observed $\sim$70\% of the sky in the $I$ and $K$ bands, and detected 5612 infrared bright sources with $m_K \lesssim 3$~mag. The TMSS was a single-epoch survey. As a logical extension of the TMSS survey, the Two Micron All Sky Survey (2MASS: \cite{skrutskie}) observed the whole sky, and provided uniform $J$, $H$,  and $K_{\rm s}$ photometric data of over 300 million sources. This 2MASS survey was also a single-epoch survey. Now the 2MASS data becomes a de facto standard in observational astronomy. However, bright sources are saturated in the 2MASS survey, and their photometry have more than 15\% errors. Figure~\ref{errorplot} is a heatmap of apparent magnitudes and photometric uncertainties of the 2MASS sources brighter than 7~mag. It shows indistinct saturation limits resulting from varying observational conditions (e.g., seeing and sky background). The figure suggests that the sources brighter than $\sim$5.5, 5, and 4.5~mag in the $J$, $H$, and $K_{\rm s}$ bands, respectively, have larger photometric errors than those fainter. Figure~\ref{errorplot2} is another expression of the fact that we astronomers do not have accurate NIR photometric data of bright sources. It is a histogram of the number of sources in the $K_{\rm s}$-band in the 2MASS (red), and in the $K$-band in the TMSS (blue) catalogs. Sources brighter than 7~mag are shown. Transparent colors show all sources regardless of the photometric error, and the opaque colors indicate sources with a photometric error smaller than 0.05~mag (corresponding to the signal to noise ratio of $\sim$20). It is clear that there is a desert of data where the accurate photometric data is absent for sources with an apparent magnitude between 2 and 4~mag. The defect is the same or even worse for the $J$ and $H$ band data, which were not observed by the TMSS survey. Also, the brighter part of the Figure~\ref{errorplot2} indicates that the TMSS survey is incomplete. \citet{kidger2003} published high precision NIR photometry of 359 bright stars ($-1 \lesssim J, H,$ or $K_{\rm s}$ [mag] $\lesssim 10$) visible from the northern hemisphere. However, this is still far from satisfactory in terms of number of stars and completeness.

There are not many large-scale infrared multi-epoch surveys. One such survey is the VISTA Variables in the Via Lactea (VVV) survey, which have observed the bulge of the Milky Way and southern disk in the NIR (\cite{minniti2010}). The VVV survey will eventually provide infrared multi-epoch data on tens of thousands of faint variables but not on those too bright for the VISTA telescope (The VVV saturation limit is typically $K_{\rm s} \sim$ 10 mag). The other is the survey with the Okayama Astrophysical Observatory Wide Field Camera\footnote{The OAOWFC has a wide field of view of 0.48$^{\circ}$ $\times$ 0.48$^{\circ}$ with a pixel scale of 1.67$^{''}$~pixel$^{-1}$.} (OAOWFC) that has been observing the Galactic plane for variability and searches for transients in the NIR (\cite{yanagisawa2014, yanagisawa2019}). The saturation limit of the OAOWFC survey is $\sim$7~mag in the $K_{\rm s}$-band (Yanagisawa, private communication). Even after the VVV and the OAOWFC surveys are completed, we will not have a uniform and systematic infrared multi-epoch data on bright stars. There is an attempt to use a local attenuation filter for monitoring bright stars (\cite{nagayama2016}), however, it only covers a small field of view and not suitable for survey-type observations.

The present observational situation of bright stars can be summarized as follows: (1) Accurate photometric data of bright stars are provided by the TMSS survey only in the $K$-band, and not in $J$ or $H$, (2) The TMSS survey is incomplete, (3) Both the TMSS and the 2MASS are single-epoch surveys. Multi-epoch observations of bright stars in the infrared have not systematically carried out yet. 

What can break the status quo? The answer is quite simple -- a NIR telescope with a small aperture. Especially, a one with a wide field of view (FOV) is preferable to carry out a survey type observation and to provide accurate NIR photometric data of bright stars by maximizing the number of the second calibration stars within a FOV. Driven by these motivations, we have developed a wide-field NIR imaging telescope. Our goals are to provide (a) high precision $J$, $H$, and $K_{\rm s}$ band photometry of bright stars, and (b) their accurate multi-epoch data with a good cadence.

The remainder of this paper is organized as follows. Section 2 describes the NIR telescope developed. Section 3 describes the performance of the infrared detector used for the telescope. Our survey designs and scientific goals are described in section 4. Then we discuss how we calibrate our photometry in section 5. An evaluation of the overall astronomical performance follows in Section 6. Finally, the summary is given in section 7.

\begin{figure}
 \begin{center}
  \includegraphics[angle=0,scale=0.7]{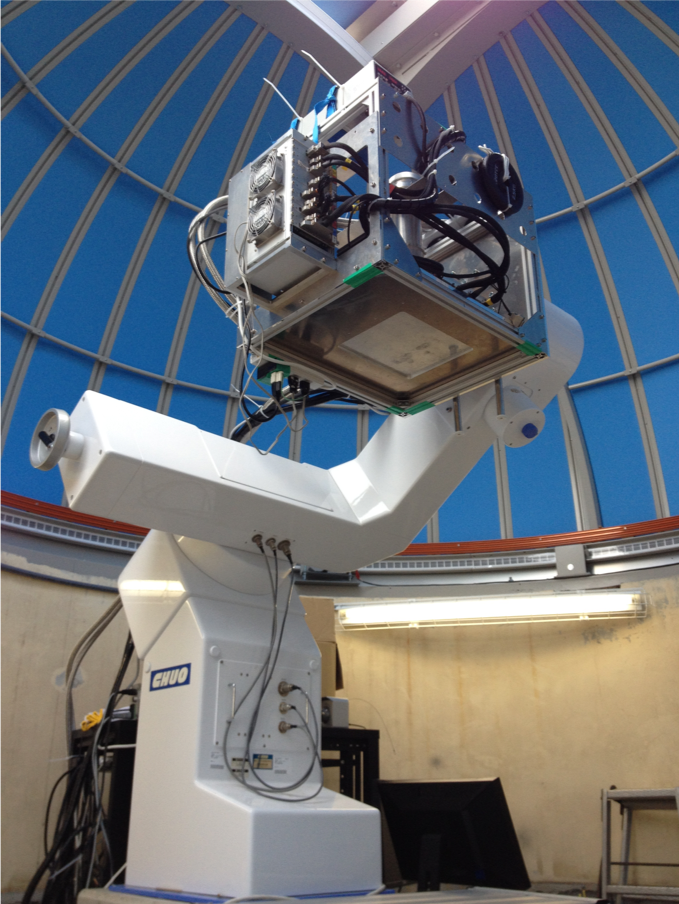} 
  \includegraphics[angle=0,scale=0.23]{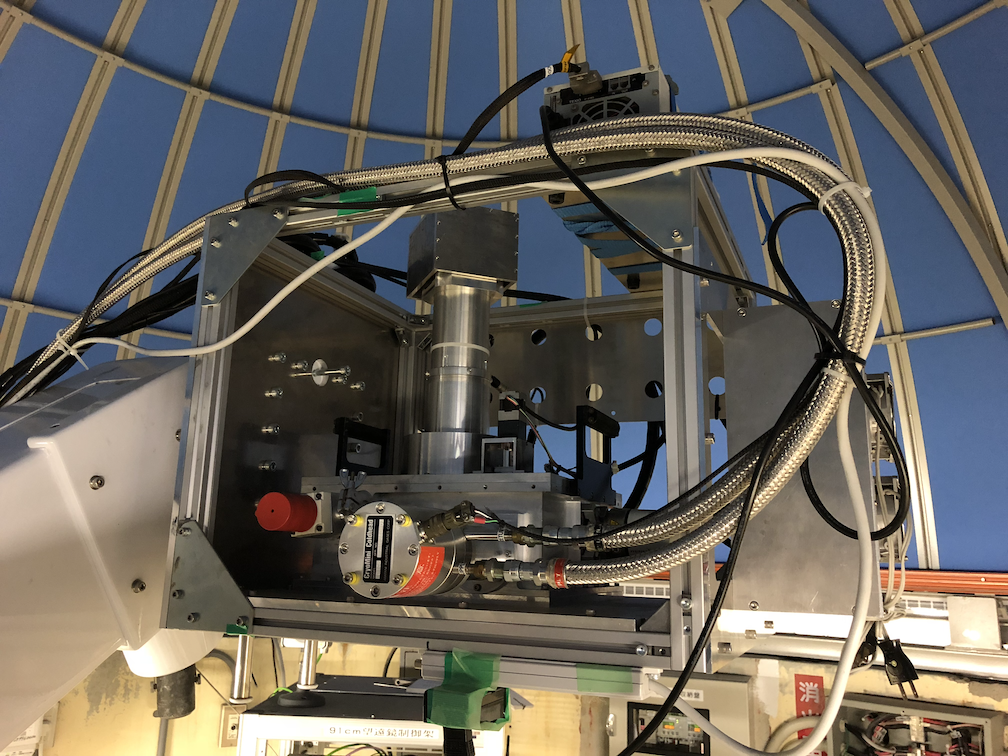} 
 \end{center}
\caption{Pictures of the TMMT observing system and the close-up of the telescope.}
\label{fig:system}
\end{figure}

\begin{figure}
 \begin{center}
  \includegraphics[angle=0,scale=0.33]{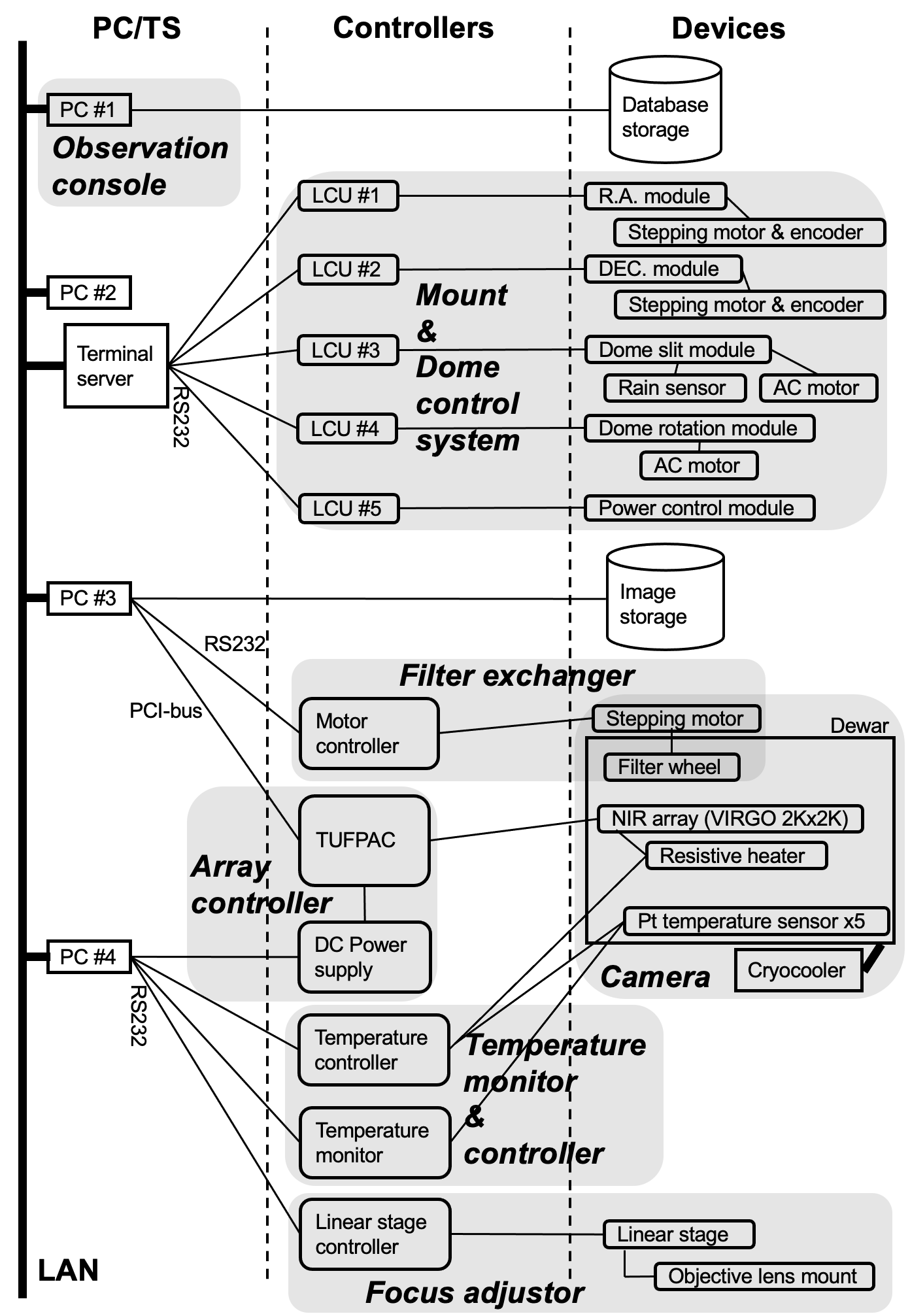} 
 \end{center}
\caption{A hardware block diagram of TMMT, which consists of seven subsystems with communication capabilities, namely, observation console, mount and dome control system, filter exchanger, array controller, camera, temperature monitor and controller, and focus adjustor. They are indicated in slanted bold texts with gray rounded rectangles. All of them are connected to the LAN through a PC or a terminal server (TS). PC~\#2 communicates with the LCU boards through the TS and also with other subsystems through the LAN. Two cylinders represent internal disk storages.}
\label{fig:systemdiagram}
\end{figure}

\section{The TMMT observing system}
\subsection{Overview}
In this paper, we use the term ``camera" as the vacuum vessel (dewar), a cryocooler, a detector and a filter wheel. An objective lens is placed on the lens mount attached in front of the camera to construct a ``telescope". Figure~\ref{fig:system} shows our observing system, which consists of the telescope, an equatorial mount, and a dome. A close-up of the telescope is also shown in the figure.

Based on the motivations mentioned in the previous section, we have developed a NIR telescope based on an engineering grade Raytheon VIRGO 2048 $\times$ 2048~pixels HgCdTe array (\cite{love2004}). The telescope has an effective aperture size of 30~mm and is named the Thirty MilliMeter Telescope\footnote{The telescope is formerly called as IR-TMT but renamed to TMMT.} (TMMT). 

The camera is cooled with a cryocooler. The array temperature is monitored with platinum (Pt) temperature sensors and maintained at 76.3~K using a temperature controller and a resistive heater. The camera has a long cylindrical ``nose" that houses a cold baffle. The diameter of the upper part of the cold baffle is 30~mm and defines the effective aperture of TMMT. At the top of the nose, an objective lens (CaF$_2$ Plano-Convex lens) mount is attached via a linear stage for focusing. TMMT and auxiliary equipment, e.g., an array controller (TUFPAC; \cite{ichikawa2003}) and a DC power supply are assembled in a bracket of aluminum frames. The total weigh of TMMT including the mounting bracket is $\sim$50~kg.

TMMT is installed on a commercial fork equatorial mount in the 4~meter dome at Subaru Telescope Okayama Branch Office (OBO; formerly known as Okayama Astrophysical Observatory), National Astronomical Observatory of Japan (NAOJ) at Okayama prefecture, Japan. The dome is located at 34.56$^\circ$ North and 134.58$^{\circ}$ East with an altitude of 372~m. The typical seeing in the optical at OBO is $\sim$1.5$^{''}$ full width at half maximum. The assemblage of TMMT, the dome, and the equatorial mount is called as the TMMT observing system hereafter. 

The TMMT observing system consists of seven main subsystems: observation console, mount and dome control system, filter exchanger, array controller, camera, temperature monitor and controller, and focus adjustor. Figure~\ref{fig:systemdiagram} is the hardware block diagram of TMMT, which lists the components of each subsystem. All the subsystems are connected to the LAN via a PC or a terminal server, so that the observation console, PC~\#1, can collect all statuses, and then issue commands as requested by observers or controllers. All of the collected statuses are stored in a database on the PC~\#1. The observing system is designed so as to be friendly for semi-automated remote observations (see section \ref{sec:operation}). 

The main specifications of the TMMT observing system are listed in Table~\ref{tab:spec}.

\begin{table}
  \tbl{Specifications of the TMMT observing system.}{
  \begin{tabular}{lrrr}
      \hline
      \hline
      Aperture Diameter [mm] & \multicolumn{3}{c}{30} \\ 
      Observation site & \multicolumn{3}{c}{N $34.56^{\circ}$, E $134.58^{\circ}$} \\
      \hline
      Detector & \multicolumn{3}{c}{Raytheon VIRGO-2K} \\
       & \multicolumn{3}{c}{(engineer grade)} \\
      Pixel size [$\mu$m] & \multicolumn{3}{c}{20 $\times$ 20} \\ 
      Array configuration [pixel] & \multicolumn{3}{c}{2048 $\times$ 2048} \\ 
      Number of output & \multicolumn{3}{c}{4} \\ 
      \hline
      Filters & $J$ & $H$ & $K_{\rm s}$ \\ 
      \hline
      Focal length [mm] & 467.81 & 470.01 & 472.78 \\
      Diffraction limit [arcsecond] & 10.35 & 13.94 & 18.10 \\
      Pixel field of view [arcsecond] & 8.86 & 8.82 & 8.76 \\
      Array field of view [degree] & 5.04 & 5.02 & 4.98 \\
	  Cut-on wavelength [$\mu$m] & 1.17 & 1.50 & 1.99 \\
	  Cut-off wavelength [$\mu$m] & 1.34 & 1.79 & 2.31 \\
	  Effective wavelength [$\mu$m] & 1.26 & 1.64 & 2.14 \\
      \hline
      \hline
  \end{tabular}}\label{tab:spec}
%\begin{tabnote}
%This is table note.
%\end{tabnote}
\end{table}

\begin{figure}
 \begin{center}
  \includegraphics[angle=0,scale=0.32]{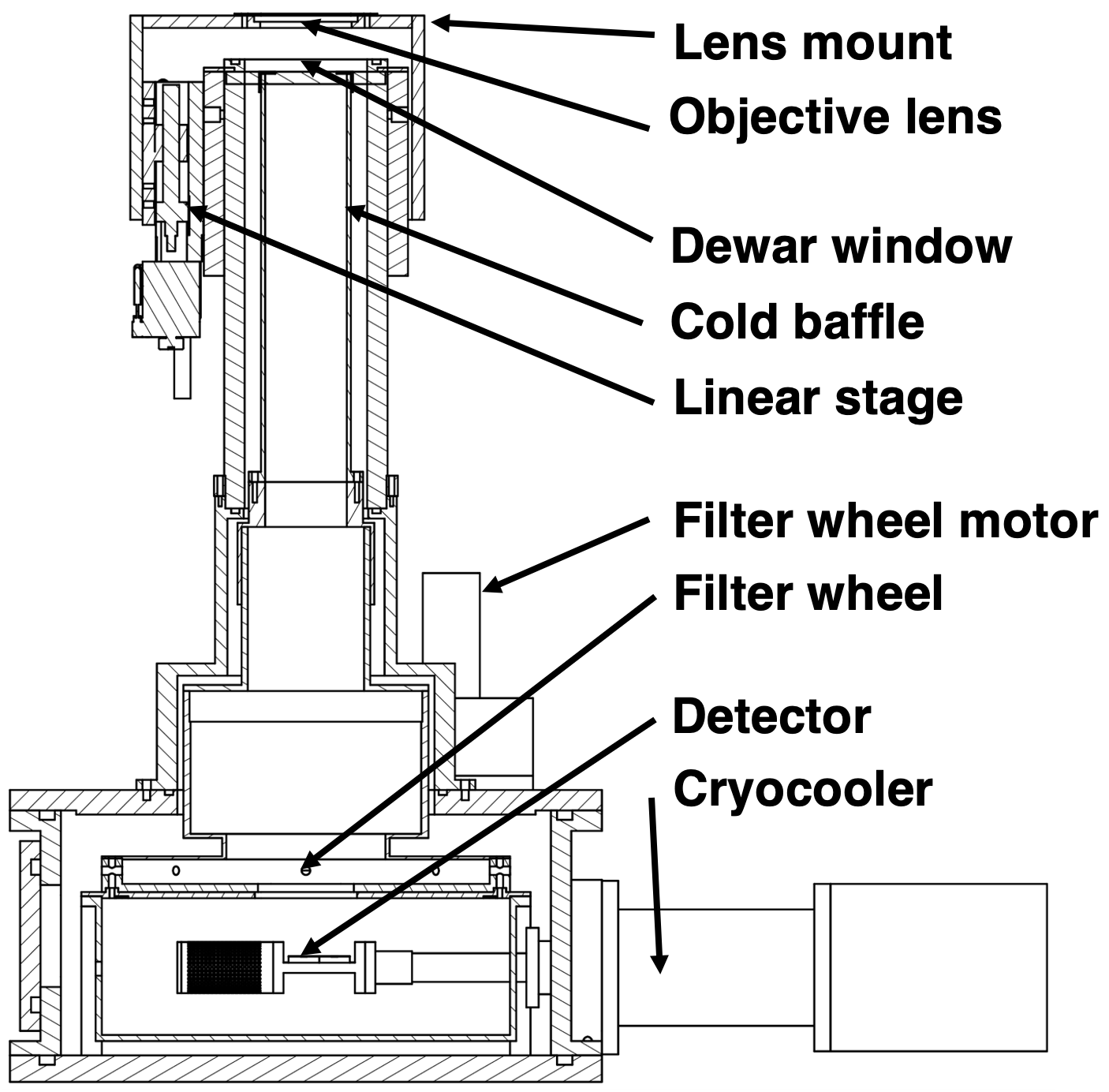} 
 \end{center}
\caption{Cutaway view of the telescope. The figure is drawn to scale. The length from the bottom of the cryostat to the lens mount is 525~mm.}
\label{fig:telescope}
\end{figure}

\subsection{Telescope}
\subsubsection{Camera}
\label{ref:camera}
Figure~\ref{fig:telescope} is a cutaway view of the telescope. The camera dewar includes a cylindrical cold baffle of 450~mm in length. The cold baffle is designed to eliminate the thermal emission from the wall of the dewar at ambient temperature, so that the main thermal background emission (except sky background) originates only from the objective lens and the dewar window. The length of the cold baffle matches the focal length of the objective lens, 472.78~mm in the $K_{\rm s}$-band. The cold baffle is connected to the cooler radiation shield of the detector and the filter wheel.

Cooling capacities of the cryocooler (ULVAC, Model D105) are 4.5~W at 77~K and 1~W at 20~K for the first and second stages, respectively. The second stage is powerful enough for keeping the detector at its operational temperature of 76.3~K. The cryocooler stably operates and keeps the detector at the operational temperature within $\pm$0.2~K for over 8~months. The detector temperature is controlled by a temperature controller (Lakeshore, model 331) with a resistive heater located next to the detector. Temperatures at the base and tip of the cold baffle, the radiation shield, the detector, and the heater are monitored by temperature monitors (RKC, MA901 and Lakeshore) using five Pt temperature sensors inside the camera. The typical operational temperatures at the base and at the tip of the cold baffle are 130 and 170 K, respectively, where the thermal emission is negligible compared to sky background.

\subsubsection{Detector and electronics}
The camera houses an engineering grade Raytheon VIRGO 2048 $\times$ 2048~pixels HgCdTe array (\cite{love2004}), which is sensitive from 0.9 to 2.5~$\mu$m. The physical pixel size is 20~$\mu$m $\times$ 20~$\mu$m. The detector is controlled by the Tohoku University Focal Plane Array Controller (TUFPAC: \cite{ichikawa2003}), which is originally designed to control the HAWAII2 array (\cite{kozlowski1998}, \cite{vural1999}). TUFPAC is operated by a PC (PC~\#3 in Figure~\ref{fig:systemdiagram}) equipped with commercially available digital signal processor boards on its PCI buses and a DC power supply (TEXIO, PW24-1.5AQ), which is controlled by another PC (PC~\#4 in Figure~\ref{fig:systemdiagram}). The original digital signal processor used in TUFPAC is obsolete and has been replaced by a commercially available bus master digital input/output interface for Linux (PCI-2772C manufactured by Interface Amita Solutions, Inc.). Due to the limitations of the current TUFPAC (e.g., the number of the I/O channels), VIRGO is operated in correlated double sampling with 4-output mode, with a minimum exposure time of 6.1~s, which is sufficiently short with no impact on our objective of observing bright stars.

\subsubsection{Telescope optics : the optical design of the telescope and its focusing system}
The optical design of the telescope is quite simple. The optics consists of only two optical components besides the filters. One is a CaF$_2$ flat plate window of the dewar, and the other is an IR Grade CaF$_2$ Plano-Convex objective lens (Edmund, \#10-63255) that is exposed to air at ambient temperature. 

The focal depth of the telescope is comparatively deep (note that the F-value of the telescope is 15.76 in the $K_{\rm s}$-band), but it is not deep enough to compensate for the axial chromatic aberration of the objective lens. Due to the chromatic aberration, the focal length varies among the filters, and therefore the pixel scale changes accordingly (see Table~\ref{tab:spec}). The differences of the focal lengths due to the chromatic aberration is up to 5~mm between $J$ and $K_{\rm s}$ bands. On the other hand, if we adopt the diameter of the airly disk as the diameter of the permissive circle of confusion, the focal depth would be $\pm$1.3~($\pm$0.75)~mm in the $K_{\rm s}$ ($J$) band. Therefore, the telescope is equipped with a focus adjustor, implemented by changing the position of the objective lens held by the lens mount attached to the motorized linear stage (SURUGA SEIKI, KXG06029-C).

\begin{figure}
 \begin{center}
  \includegraphics[angle=0,scale=0.3]{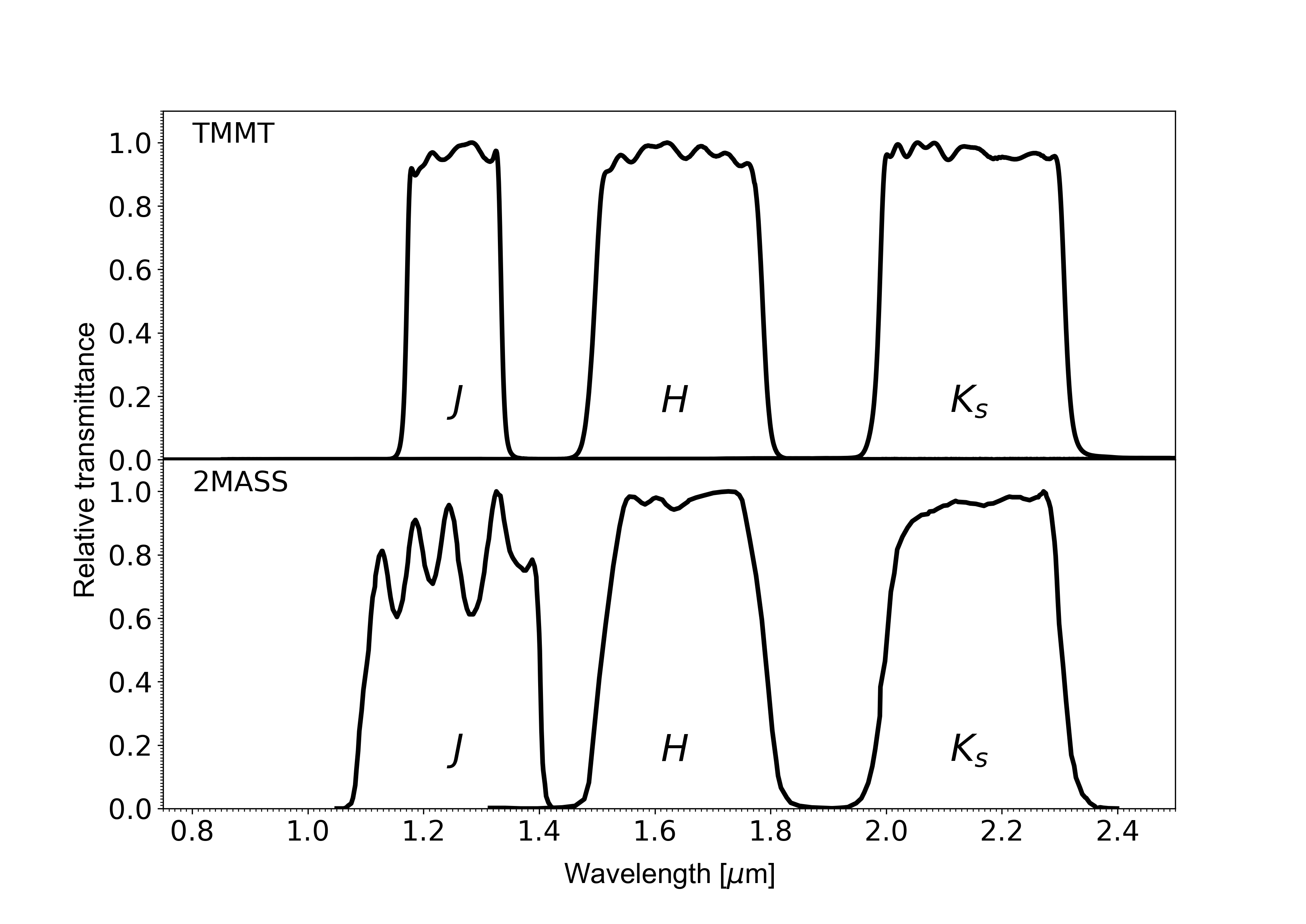} 
 \end{center}
\caption{Normalized relative transmittances of $J$, $H$, and $K_{\rm s}$ filters for TMMT (upper) and 2MASS (lower).}
\label{fig:trans}
\end{figure}

\subsubsection{Filters : filter exchanger and filter system}
A filter wheel is located at the cryo temperature inside the dewar and is directly driven by a stepping motor (ORIENTAL MOTOR, PK523A-M30) attached outside the camera. Up to four filters can be mounted on the filter wheel. In addition to a metal plate for taking dark images, Mauna Kea Observatories Near-Infrared (MKO-NIR) $J, H$, and $K_{\rm s}$ filters (\cite{simons2002}, \cite{tokunaga2002}, \cite{tokunaga2005}) are loaded. The filters are circular in shape, with a physical size of 60~mm in diameter and 3.5~mm in thickness. A clear aperture of the filter frame is 55~mm in diameter, and the filters are located 23~mm from the detector. The filters are large enough, so that the vignetting is negligible over the field of view.

The transmittance of each filter at liquid nitrogen temperature has been measured using a spectrophotometer (Shimadzu UV-3100) at the Advanced Technology Center, NAOJ. At least five measurements are made for each filter, and the median value of the measurements is adopted. Figure~\ref{fig:trans} shows the relative transmittance (the maximum value normalized to one) of each filter along with those of the 2MASS filters for comparison. Based on these transmission curves, we calculate the cut-on and cut-off wavelengths and the ``effective wavelength" of the three filters. The cut-on and cut-off wavelengths are defined as wavelengths where the relative transmittance drops to 0.5, and the ``effective wavelength" is defined as 
\begin{equation}
\lambda_0 = \frac{\int{\lambda R_\lambda} d\lambda}{\int{R_\lambda} d\lambda},
\end{equation}
where $R_\lambda$ is the relative transmittance of the filter. Note that the definition does not take into account atmospheric transmission, throughput of the telescope (e.g., CaF$_2$ lenses), and quantum efficiency of the detector. The results are summarized in Table~\ref{tab:spec}.

% http://www.dppobservatory.net/DomeAutomation/dome_synchronisation.pdf
\subsection{Mount and dome control system}
The equatorial fork mount (R.A. and DEC. axes), dome rotation, dome slit, and power supplies have in-house Local Control Unit (LCU\footnote{LCU is a microcomputer based parallel I/O board with a serial interface, that is used to control subordinate devices and to make the subsystem a discrete controller.}) boards, and are controlled from a PC (PC~\#2 in Figure~\ref{fig:systemdiagram}) using the RS232C protocol through a terminal server\footnote{Terminal server is a Linux based appliance used to convert messages from serial to TCP/IP or vice versa.}. In case of emergency, such as a power or network failure, the power supplies can be automatically shut off. It also monitors the rain sensor and automatically closes the dome slit when rain is detected. Thanks to this system, semi-automatic remote observations can be done safely. The specifications of the mount and dome, as well as the details of the control system, are found in \citet{tsutsui2016}.

%\citet{tsutsui2016} developed a control system for the 4-meter dome and the mount, which controls the equatorial fork mount (R.A. and DEC. axes), dome rotation, dome slit, and power supplies. Each of these device modules has an in-house Local Control Unit (LCU\footnote{LCU is a microcomputer based parallel I/O board with a serial interface, that is used to control subordinate devices and to make the subsystem a discrete controller.}) board. The LCU boards communicate with a PC (PC~\#2 in Figure~\ref{fig:systemdiagram}) using the RS232C protocol through a terminal server\footnote{Terminal server is a Linux based appliance used to convert messages from serial to TCP/IP or vice versa.}. 

\subsection{Operation}
\label{sec:operation}
We have developed a software that links the subsystems together to enable semi-automatic remote observations. The software simply receives the list of target coordinates. Then the software sequentially repeats target pointing, exposure, dithering, filter exchange, etc. until the twilight time. Targets closer to the western horizon will be given priority to maximize the cadence of the observations. Since we can leave the observation to the software, survey-type observation can be easily done. It also requires less manpower, so that nightly observations can be continued over a long period of time.

\section{Detector performance}
Since the detector used for TMMT is an engineering grade VIRGO, its performance should be evaluated and properly calibrated before any scientific use. In this section we discuss its pixel operability, dark current, linearity, conversion gain, readout noise, and relative response of the pixels. First, we describe the data used for the following analyses.

\subsection{Data}
\subsubsection{Dark}
\label{sec:darkdata}
The temperature of the detector was controlled at 76.3~K. A series of dark images were taken with an exposure time of 7~s and from 10~s to 240~s by 10~s step. Ten dark images were taken for each exposure time.

\subsubsection{Dome ceiling images}
\label{sec:gaindata}
The dome ceiling, which can be regarded as a pseudo-constant source of light, was observed in the $K_{\rm s}$-band with exposure time from 7 to 200~s. Five images were taken for each exposure time. The intensity of the thermal radiation from the dome ceiling depends on the outdoor air temperature. Expecting small variation of the temperature, we took the data on a cloudy night. The average outdoor temperature during the night was $\sim$10~$^\circ$C, and the change was within $\pm$2~$^\circ$C. The signal count of the dome ceiling in the $K_{\rm s}$-band was roughly 300~ADU~s$^{-1}$~pixel$^{-1}$. This corresponds to 7.8~mag~pixel$^{-1}$. To properly calibrate the temporal variation of the thermal radiation from the dome ceiling, we monitored the temperature change by taking a 7~s exposure image immediately after each exposure.

\subsubsection{Twilight sky images}
\label{sec:flatdata}
The relative response of the pixels can be calibrated by imaging an evenly illuminated light source. For this purpose, we took twilight sky images for each filter at the beginning and end of each night. It takes several nights to obtain the images for all filters. In order to achieve better signal to noise ratio, we accumulate the twilight images as many as possible on nights in good sky conditions.

\subsection{Pixel operability}
\begin{figure}
 \begin{center}
  \includegraphics[angle=0,scale=0.3]{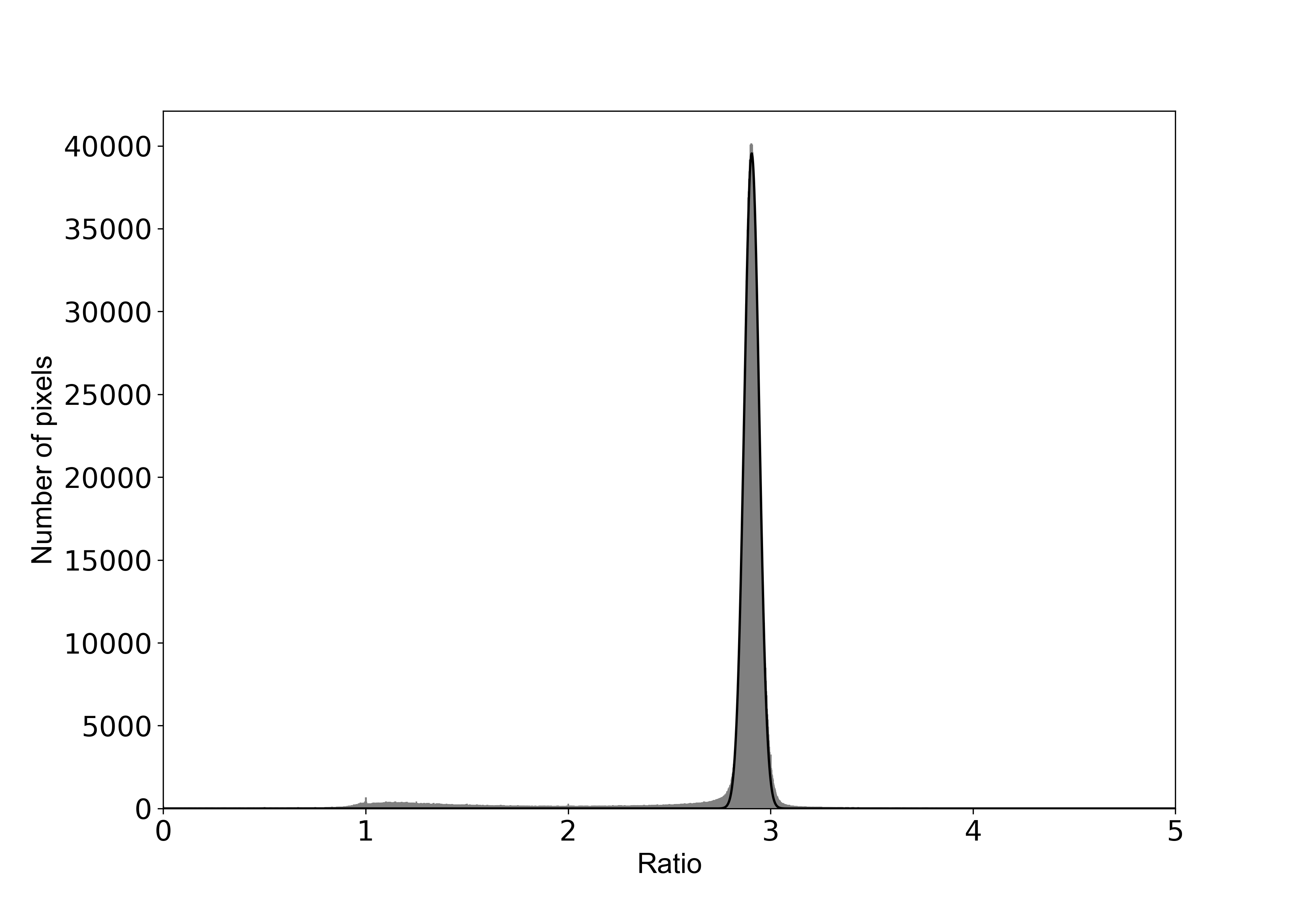} 
 \end{center}
\caption{Histogram of pixel values for an image that is made by dividing the 60~s exposure dome ceiling image with that of 20~s exposure. The width of the bin is set to 0.001. The thick solid line shows the best fit Gaussian to the distribution.}
\label{fig:ratio}
\end{figure}

\begin{figure}
 \begin{center}
  \includegraphics[angle=0,scale=0.42]{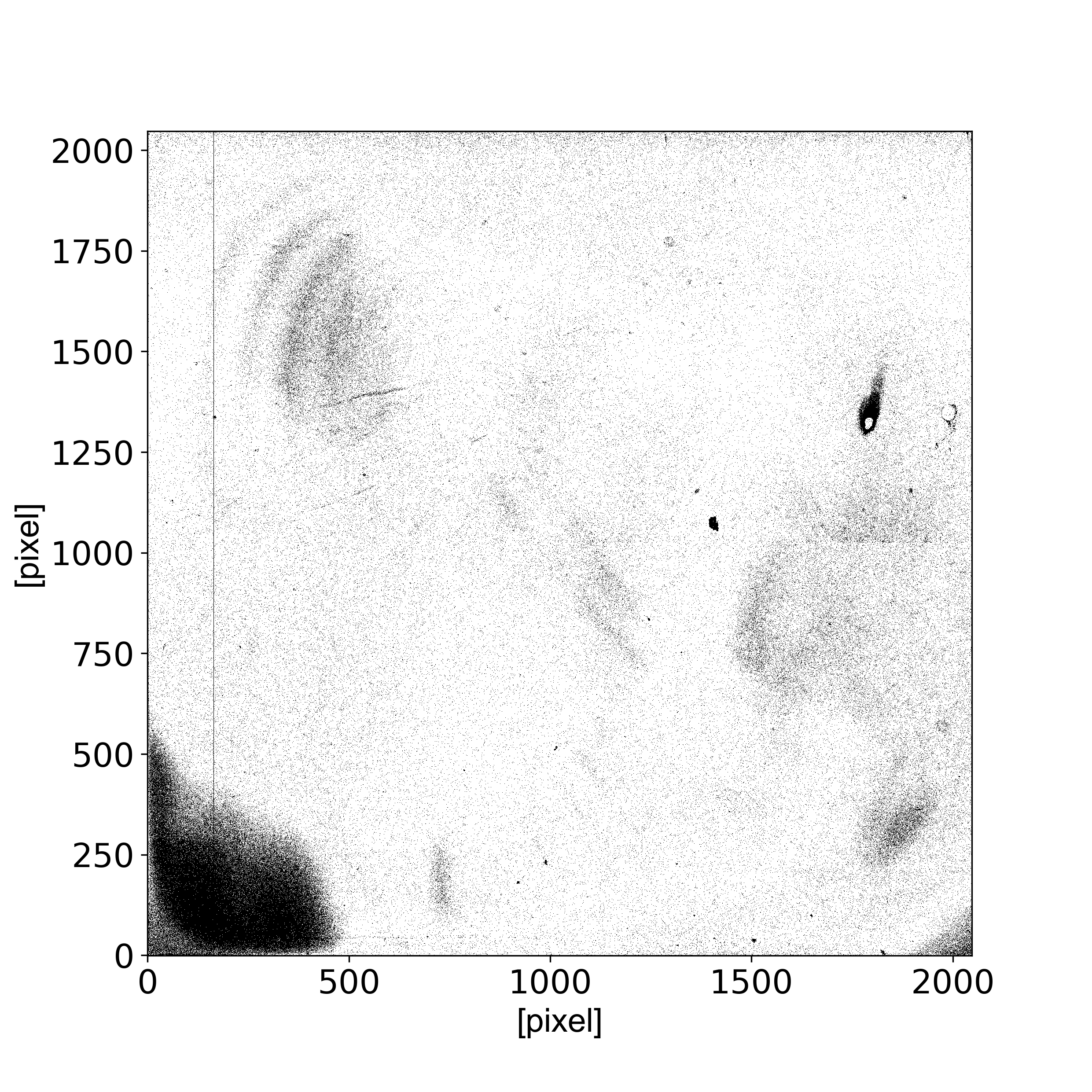} 
 \end{center}
\caption{Inoperable pixel (black) map of the detector.}
\label{fig:operablemap}
\end{figure}

Here we define an inoperable pixel as the one that produces no useful signal, namely, bad, dead, and hot pixel, that has poor sensitivity, poor linearity, and/or that is located in seriously vignetted region. The number of inoperable pixels was counted as follows.

First, an image was made by dividing the dome ceiling image of 60~s exposure with that of 20~s exposure. No correction was made for the temporal variation due to the change of thermal radiation from the dome ceiling. The charges accumulated in the 60 and 20~s exposures are $\sim$60 and $\sim$20\% of the full-well of the detector, respectively. Ideally, the signal counts of every pixel in this image should be equal to the ratio of the exposure times, which is 3 in this case. Figure~\ref{fig:ratio} is a histogram of the signal counts of every pixel in the image. The distribution has a peak at 2.91 with the standard deviation $\sigma$ of the distribution of 0.04. Then we refer to pixels whose values deviate from the peak value by more than 5~$\sigma$ as group A of inoperable pixels. Obviously, the histogram is affected by the nonlinearity of the detector and/or the temporal variation of the thermal radiation from the dome ceiling. In addition, the dark current is ignored for this study. However, these factors are not important for our purpose to identify inoperable pixels.

Next, ten dark images with an exposure time of 20~s are combined to obtain the median image. The median and standard deviation of the pixel values of the median combined image are calculated. Then we choose pixels whose values deviate from the median by more than 5~$\sigma$ as group B of inoperable pixels.

Finally, a disjunction operation (i.e. logical OR) between candidate of groups A and B is performed to identify final inoperable pixels. The number of inoperable pixels is 396,835. As a result, the pixel operability of the detector is approximately 90.5\%. This is consistent with a report by Raytheon Corporation, in which 92 $\sim$ 94\% of the pixels passed their quantum efficiency test. Figure~\ref{fig:operablemap} shows the map of inoperable pixels identified by our study. The identified inoperable pixels will be masked and not be used for further analyses.

\subsection{Gain and readout noise}
\begin{figure}
 \begin{center}
  \includegraphics[angle=0,scale=0.4]{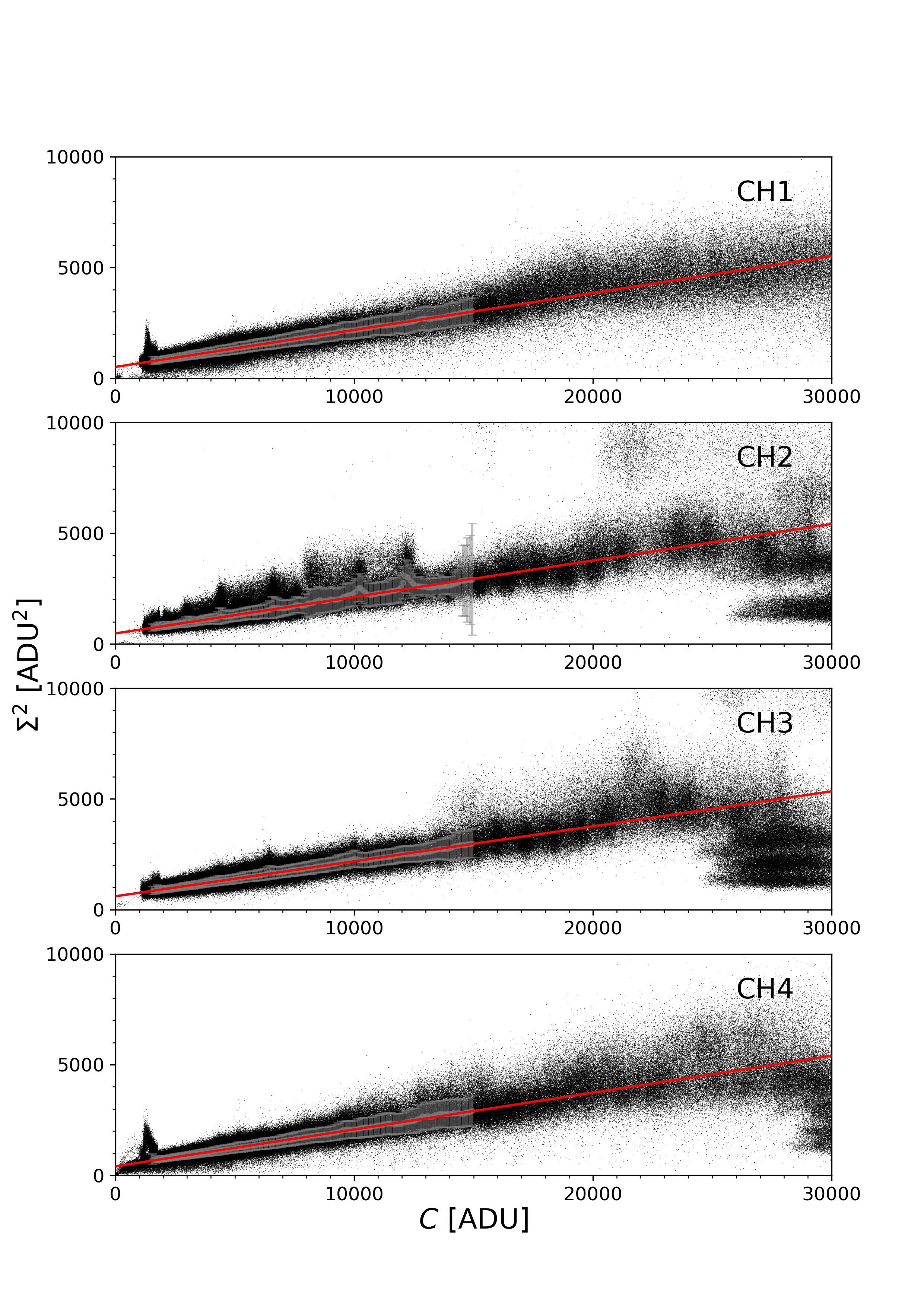} 
 \end{center}
\caption{The half variance of difference images vs. the signal count of images for channels 1 to 4 from the top to the bottom. White points with error bars show the running median and red lines show their fit.} 
\label{fig:gain}
\end{figure}

\begin{table}
  \tbl{Conversion gain, readout noise, and the dark current generation rate of each channel.}{
  \begin{tabular}{lrrr}
      \hline
      Channel & Gain [e$^-$~ADU$^{-1}$] & Readout noise [e$^-$] & Dark current [e$^-$~s$^{-1}$] \\ 
      \hline
      ch1 & 5.99 $\pm$ 0.01 & 14.20 $\pm$ 0.08  & 0.38 $\pm$ 0.02 \\
      ch2 & 6.07 $\pm$ 0.05 & 12.93 $\pm$ 0.32  & 0.37 $\pm$ 0.03 \\
      ch3 & 6.33 $\pm$ 0.02 & 15.12 $\pm$ 0.15  & 0.33 $\pm$ 0.02 \\
      ch4 & 6.01 $\pm$ 0.02 & 11.38 $\pm$ 0.11  & 0.57 $\pm$ 0.03 \\
      \hline
  \end{tabular}}\label{tab:gain}
%\begin{tabnote}
%This is table note.
%\end{tabnote}
\end{table}

The conversion gain (e$^-$~ADU$^{-1}$) relates the raw detector unit of analog-to-digital unit (ADU) to the electron unit (e$^-$). Because we employ the 4-channel readout of the VIRGO detector with a vertical strip of 512 $\times$ 2048~pixels for each channel, the gain and the readout noise are derived for each channel. For convenience, we define the N-th channel as an image section of 512 $\times$ (N-1) to 512 $\times$ N-1 pixels in the column direction and 0 to 2047 pixels in the row direction.

We use a photon transfer curve method to obtain the conversion gain $g$ and the readout noise $R$ in e$^-$ of the detector. The method assumes that number of photons detected by the detector follows the Poisson statistics, and that all the pixels used for the measurement have same properties and same illumination. Then, $g$ and $R$ are related by
\begin{equation}
\left( g \times \Sigma \right)^2 = g \times C + R^2,
\end{equation}
where $C$ is the count of pixel value in ADU and $\Sigma$ the standard deviation. 

The $g$ and $R$ are calculated using the data described in section~\ref{sec:gaindata} after calibrating the temporal variation of the thermal radiation from the dome ceiling. A pair of the dome ceiling images with the same exposure time is subtracted from each other, and the median and variance are calculated with an iterative 3-sigma clipping for every 8 $\times$ 8~pixels bin using the ceiling images and the subtracted images, respectively. The size of the bin of 8 $\times$ 8~pixels is chosen to ensure both statistical sufficiency and uniform illumination. Then half of the variance corresponds to the $\Sigma^2$ and the median to $C$, whose plots are shown in Figure~\ref{fig:gain}. The running median of the variance with a step of 100~ADU in signal is calculated and a linear relation is fitted using data with signal values larger than 1500~ADU and smaller than 15000~ADU, where the linearity of the detector is relatively good (see section~\ref{sec:linearity}). The results of the fits and the calculated gain and readout noise for each channel are summarized in Table~\ref{tab:gain}. 

\subsection{Dark current}
\label{sec:dark}
\begin{figure}
 \begin{center}
  \includegraphics[angle=0,scale=0.29]{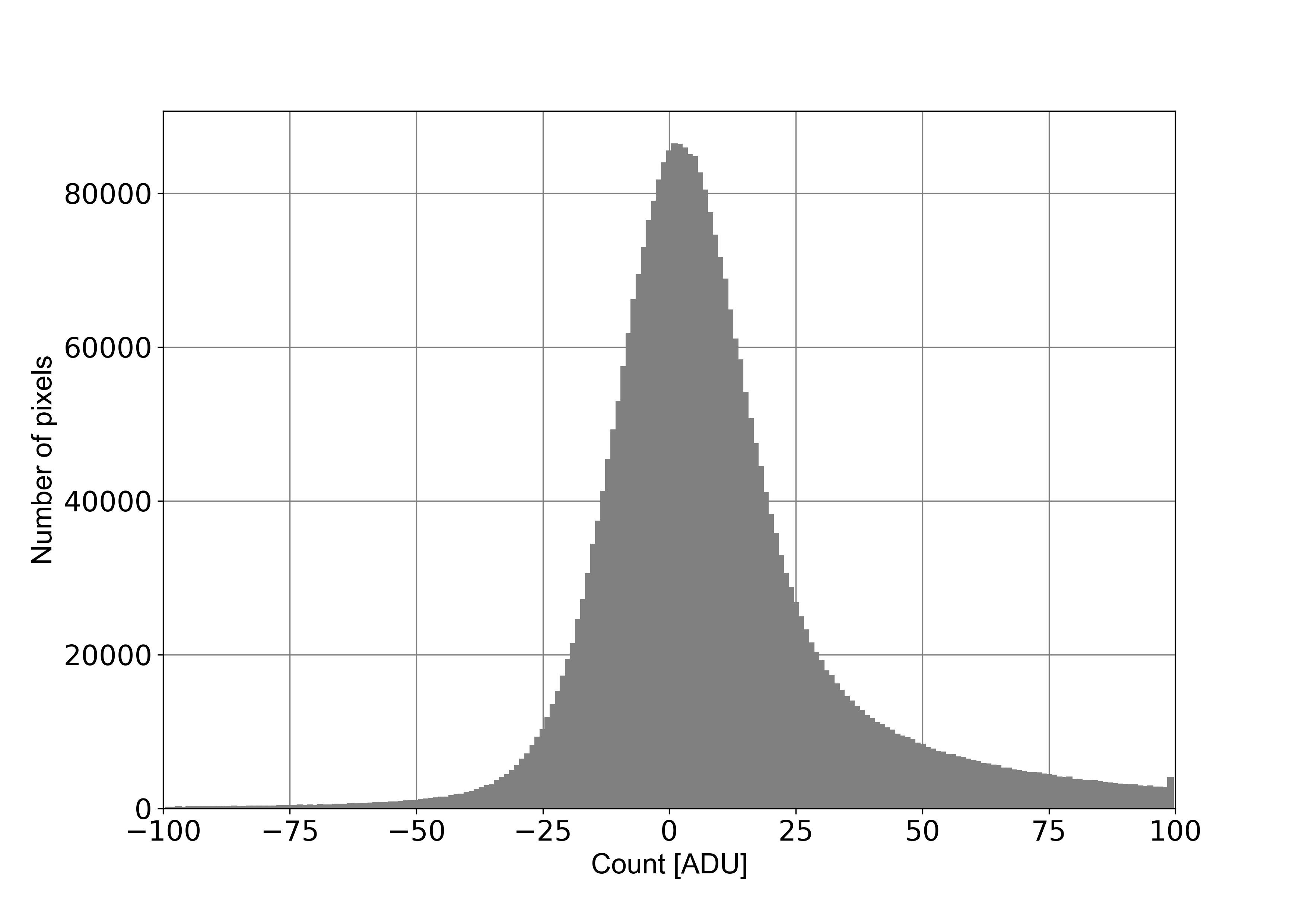} 
 \end{center}
\caption{Distribution of dark counts in a dark image that is the median combination of ten dark images of 20~s exposure. The size of the bin is chosen to be 1~ADU.}
\label{fig:dark_dist}
\end{figure}

\begin{figure}
 \begin{center}
  \includegraphics[angle=0,scale=0.4]{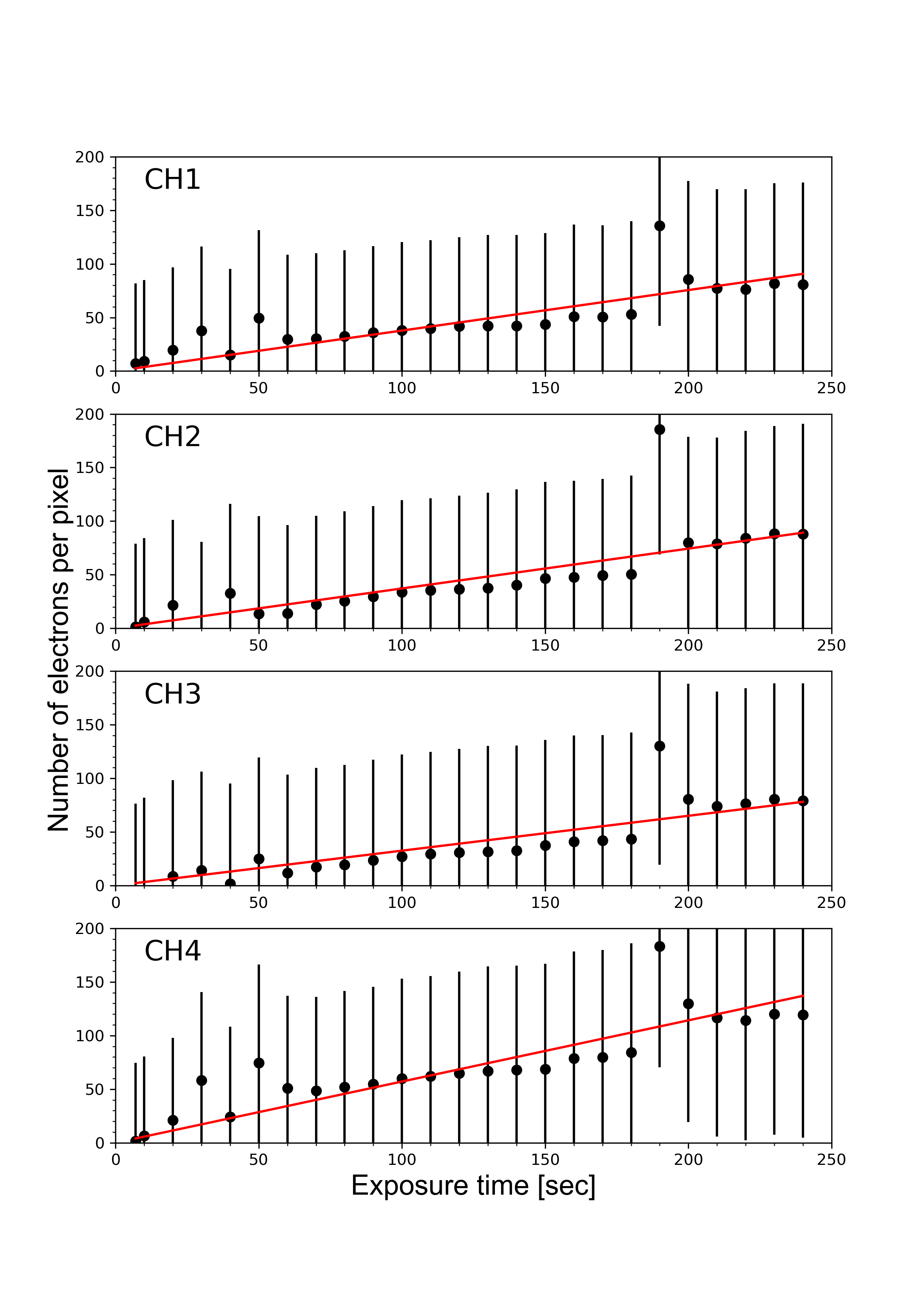} 
 \end{center}
\caption{Number of dark electrons generated is plotted against the exposure time for channels 1 to 4 from top to bottom. The solid lines show the best fit linear relation to the data. Error bars indicate the standard deviation of the fitted normal distribution.}
\label{fig:dark}
\end{figure}

The dark current generation rate is estimated using the data described in section~\ref{sec:darkdata}. For each median-combined dark image, mode and sigma of the count distribution (see Figure~\ref{fig:dark_dist} for example) are estimated by fitting a normal distribution function to the distribution to obtain representative values of dark current and its scatter. The dark current generation rate is estimated from the relationship between the exposure time and the number of electrons generated. Figure~\ref{fig:dark} shows the result, where a linear relation with a fixed intercept of zero is fitted. Its slope corresponds to the dark current generation rates, which are 0.3-0.5 ~e$^{-}$~s$^{-1}$ as summarized in Table~\ref{tab:gain}. The detector data report provided by Raytheon Corporation gives a dark generation rate of 1.51~$\pm$~1.54~e$^{-}$~s$^{-1}$ at an operating temperature of 78~K. The possible cause of the difference is the difference in the temperature, where we operate the detector at 76.3~K (see section \ref{ref:camera}).

\subsection{Linearity}
\label{sec:linearity}
\begin{figure}
 \begin{center}
  \includegraphics[angle=0,scale=0.39]{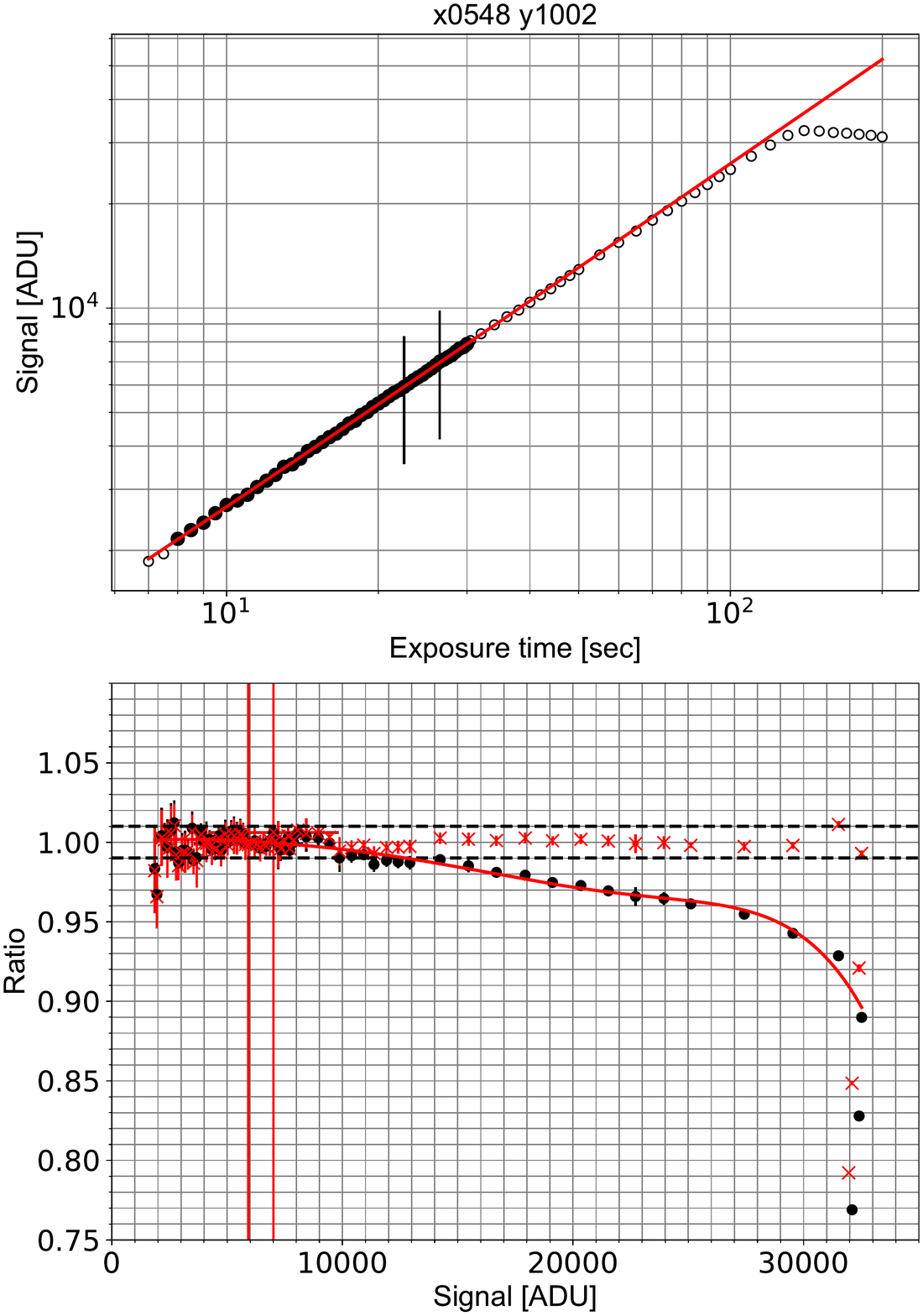} 
 \end{center}
\caption{Top panel: Exposure time vs. signal counts in ADU for a pixel ($x=$548, $y=$1002). Open circles indicate the signal counts. The error bars in the signal counts are barely visible and are smaller than the size of the circles. The thick solid line shows the best fit linear relation to the signal counts with $2000 < {\rm Signal [ADU]} < 8000$, indicated by filled circles. Bottom panel: Signal counts in ADU vs. ratios (the signal counts over the best fit values) for the same pixel. Filled circles and error bars show the ratios and their error, respectively. The thick line is the best fit of the sixth-order polynomials to the filled circles, and the crosses show the ratios after the linearity correction. The dashed lines show the $\pm$1\% deviation from the best fit linear relation.}
\label{fig:linearity}
\end{figure}

\begin{figure}
 \begin{center}
  \includegraphics[angle=0,scale=0.39]{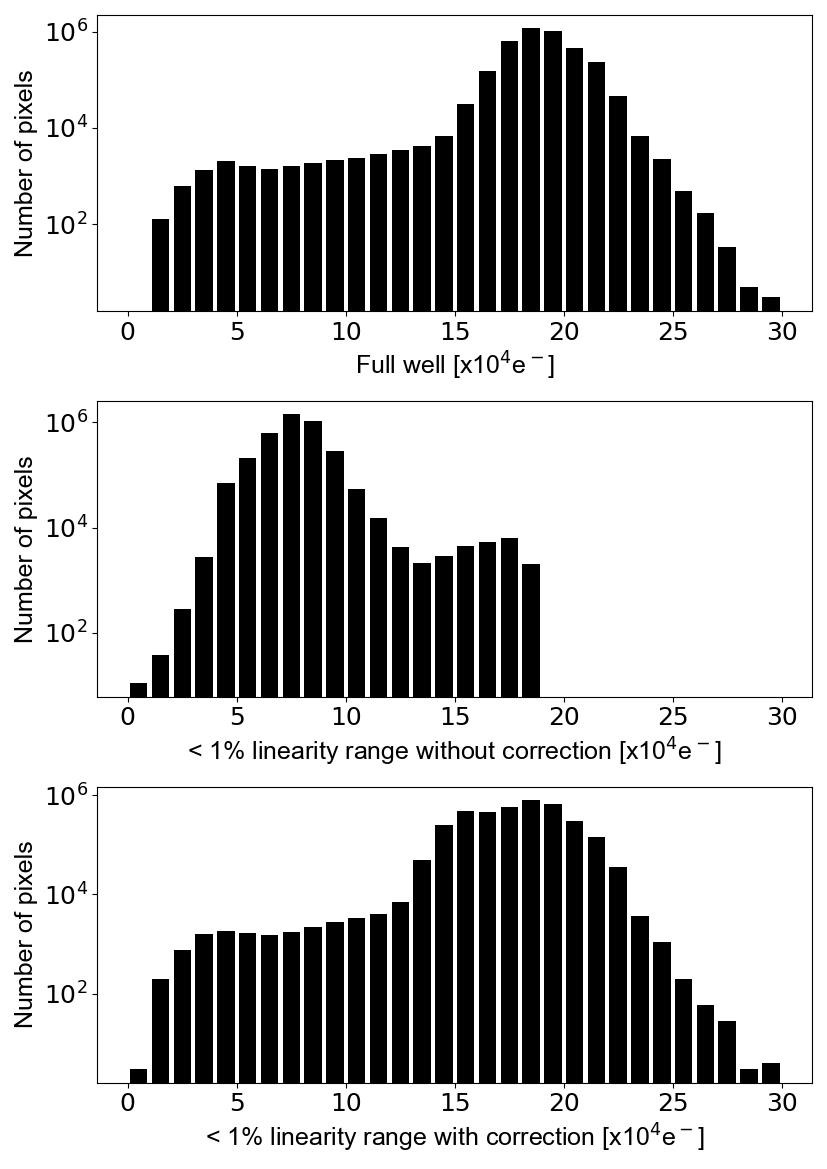} 
 \end{center}
\caption{Histograms of the full well (top) and $<$1\% linearity range without (middle) and with (bottom) the non-linearity correction for all operative pixels. The size of the bin is chosen to be 10$^6$ e$^-$.} 
\label{fig:linearityrange}
\end{figure}

\begin{table}
  \tbl{Well depths of the detector. All are in the unit of 10$^4$e$^-$.}{
  \begin{tabular}{lr}
	\hline
    Full well & 18.91$\pm$1.55 \\ 
	%\multicolumn{2}{c}{{\bf $<$1\% linearity range}} \\ 
	$<$1\% linearity range without correction & 7.72$\pm$1.31 \\ 
	$<$1\% linearity range with correction & 18.08$\pm$2.10 \\ 
    \hline
  \end{tabular}}\label{tab:linearity}
\end{table}

The linearity of the detector is evaluated using the data described in section~\ref{sec:gaindata}. First, we fit a linear relation to  the signal counts in the range of 2000 to 8000~ADU. The top panel of Figure~\ref{fig:linearity} shows the result of the regression for a pixel ($x=$548, $y=$1002) as an example. Dots in the bottom panel of Figure~\ref{fig:linearity} indicate deviation from the linear regression as ratio. Finally, a sixth-order polynomial curve is fitted to the signal count vs. ratio data. The correction for linearity can be made by dividing the signal count by the best fit prediction ratio. The results of the polynomial regression and the linearity correction are shown as thick solid line and crosses in the bottom panel of Figure~\ref{fig:linearity}. We define the applicable range of the linearity correction such that the ratio between the predicted and corrected signal counts fall within $\pm$1\% from unity. The series of analysis is performed for all operable pixels to obtain the linearity correction of sixth-order polynomial equation and its range of application. Signal counts outside the range are masked in practical data analysis. With a few exceptions, the properties of linearity are more or less the same for operable pixels. The median and standard deviation for all operative pixels of the full well and the $<$1\% linearity range without and with the non-linearity correction are summarized in Table~\ref{tab:linearity} and their histograms are shown in Figure~\ref{fig:linearityrange}.

\begin{figure}
\begin{center}
\includegraphics[scale=0.42,angle=0]{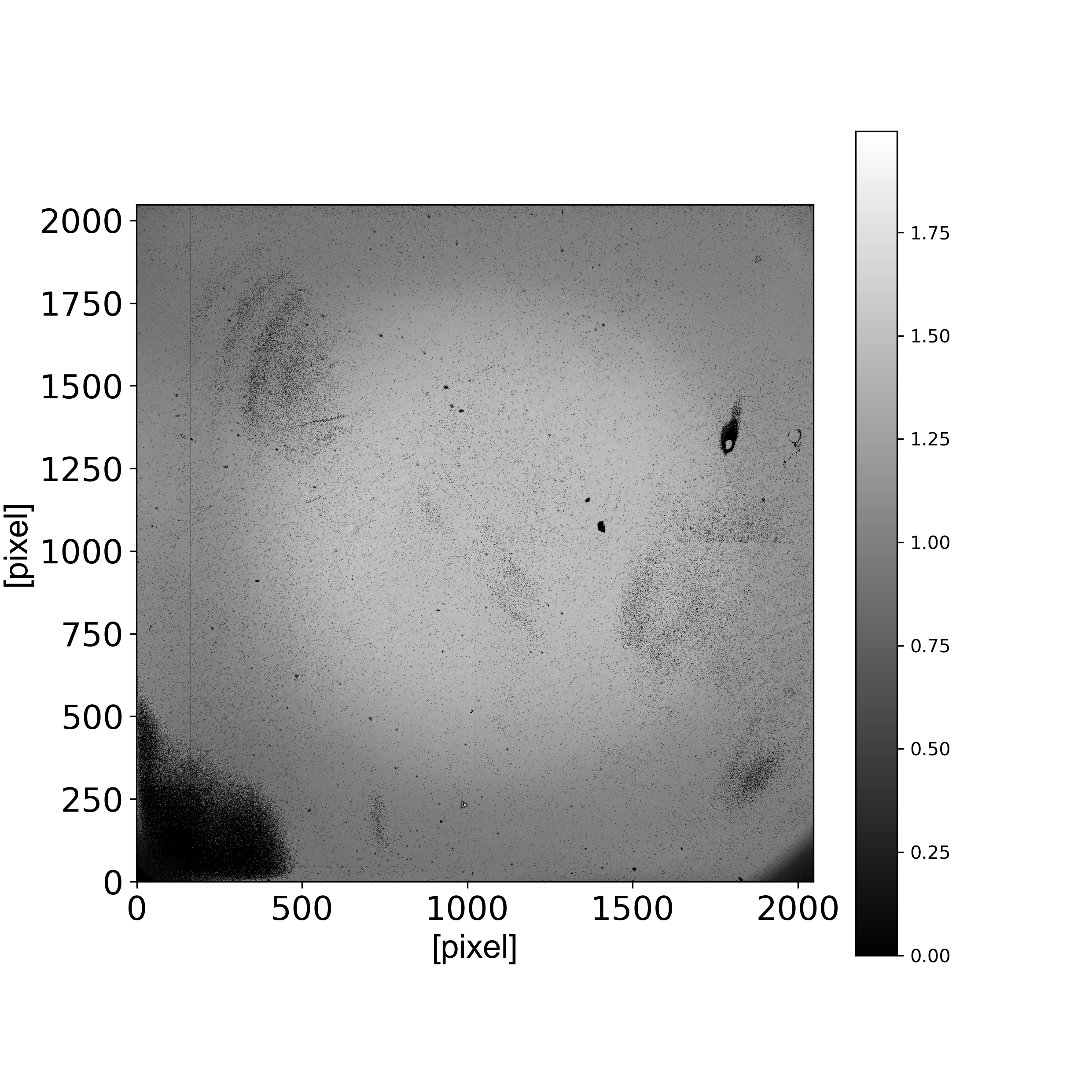} \includegraphics[scale=0.3,angle=0]{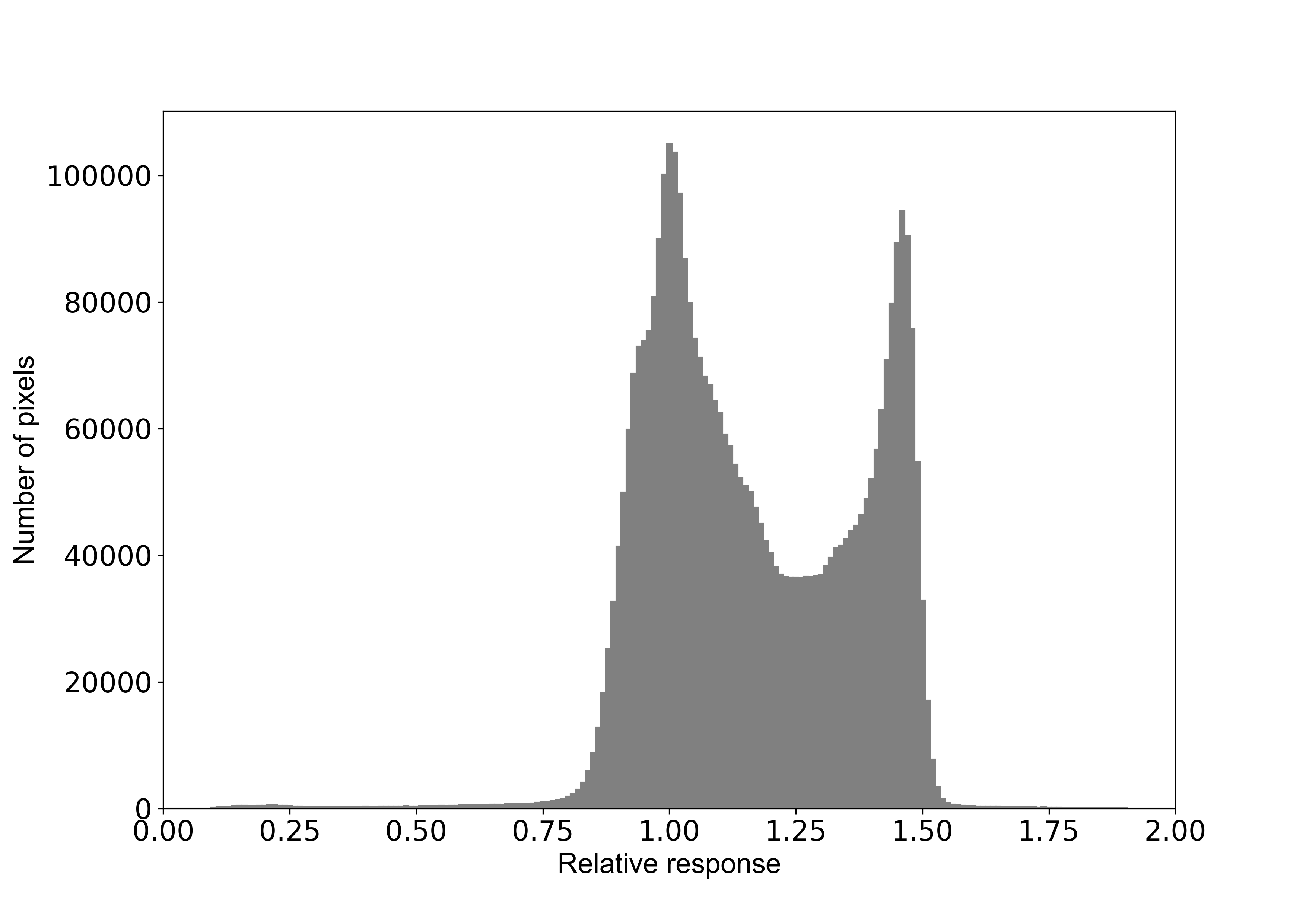} 
\end{center}
\caption{$K_{{\rm s}}$-band flat-fielding image and the histogram of its pixel values.}
\label{fig:response}
\end{figure}

\subsection{Relative response of the pixels}
More than a hundred twilight sky images (corrected for linearity and dark) are median-combined and are normalized with respect to the median value of the whole pixels. So-called flat-fielding image is made in this way. 

Figure~\ref{fig:response} shows the $K_{{\rm s}}$-band flat-fielding image and its histogram. There is a non-uniform illumination pattern in the flat-fielding image due to vignetting of the incoming light caused by the cold baffle. Note that the physical array size is 40.96~mm by 40.96~mm and the diameter of the cold baffle is 30~mm. This is the cause of the double-peaked distribution of pixel values seen in the bottom panel of Figure~\ref{fig:response}.

\section{Survey designs and scientific goals}
\label{sec:survey}
So far, we have described the design and specifications of the TMMT observing system. In this section, we describe the scientific goals of the TMMT project, which consists of two major survey observations. One is the photometric survey, and the other is the variable star monitoring survey.

\begin{figure}
\begin{center}
\includegraphics[scale=0.36]{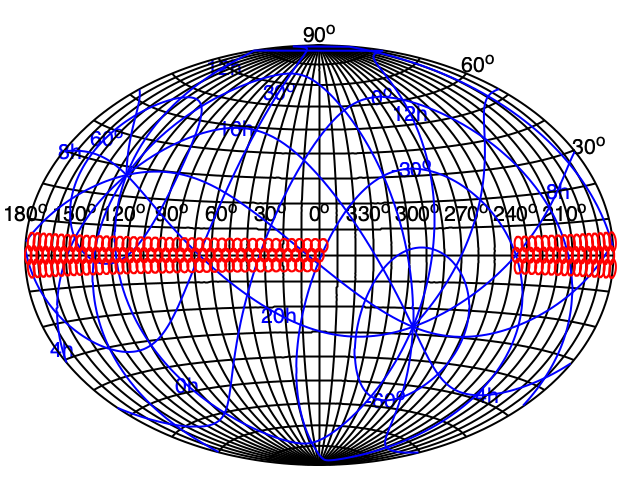} 
\end{center}
\caption{Map of the TMMT survey area in the aitoff projection in galactic coordinates. The blue lines show equatorial coordinates, and each of the red circles with a diameter of 5$^{\circ}$ represents the one FOV of TMMT.}
\label{fig:survey}
\end{figure}

\subsection{Northern sky photometric survey}
The primary purpose of the TMMT development is to observe northern bright stars in the $J$, $H$, and $K_{\rm s}$ bands, and to provide their accurate photometric data. To achieve this goal, we have carried out a survey of the belt-like area along the northern Galactic plane. Specifically, the survey area ranges within declination of $\delta \ge -30^\circ$ and galactic latitude of $|b| \le 5^\circ$ as illustrated in Figure~\ref{fig:survey}. We use a hexagonal mosaicking pattern with a spacing of $\frac{\sqrt{3}}{2} \times 5^{\circ}$ between the adjacent pointings. Thanks to the wide FOV ($\sim$ 25~square~degrees) of TMMT, the whole survey area can be covered by 173~pointings. We have carried out 20-dithering observation at each pointing for the photometric survey, where dithering positions are located on a circumference with a radius of 40$^{'}$ with an equal spacing. A 7~s exposure image is taken at each dithering position, and the total exposure time is 140~s.

\begin{figure*}
\begin{center}
\includegraphics[scale=0.46]{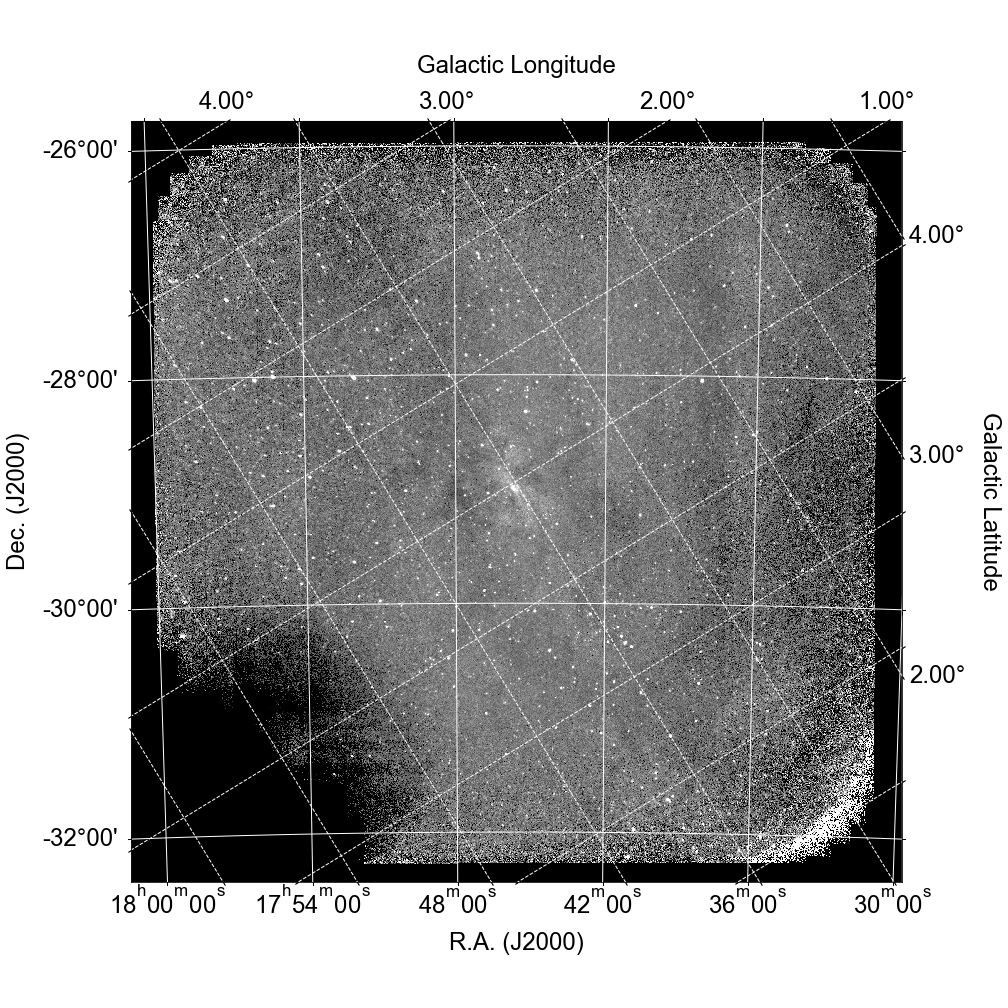} 
\end{center}
\caption{A $K_{\rm s}$-band image of Galactic center taken with TMMT. The image is made with a median combination of 20~dithered images. Solid lines indicate equatorial coordinates and dotted lines show galactic coordinates.}
\label{fig:galcen}
\end{figure*}

As an example of the image taken with TMMT, we show a $K_{\rm s}$-band image toward the Galactic center in Figure~\ref{fig:galcen}. Note that size of the image is slightly larger than the FOV of TMMT (5$^\circ$ $\times$ 5$^\circ$), as it is the result of a median combination of 20~dithered images. The prominent black area in the lower left is caused by inoperable pixels (see Figure~\ref{fig:operablemap}). We perform the  photometry on the combined image referring to its exposure map. To achieve high photometric accuracy, we have chosen the stars in the area with 10 or more exposures in their exposure map.

The photometric survey has been completed. More than 13$\times10^5$, 29$\times10^5$, and 19$\times10^5$ stars have been detected in the $J, H$, and $K_{\rm s}$ bands, respectively (Ita et al., in preparation). We plan to gradually expand the survey area in the future.

\subsection{Galactic plane variable star survey}
The secondary purpose of the TMMT observation is to characterize infrared variability of bright stars in the Galactic plane. It is known that some types of variable stars, such as Cepheids and Miras, have a statistical relationship between their brightness and light variation periods. The relation is called the period-luminosity relation (\cite{leavitt1912}) and is widely used as a good distance scale (e.g., \cite{freedman2013}). The infrared period-luminosity relation is preferable, due to smaller effects of interstellar extinction. Also, the scatter of the relation becomes smaller at longer wavelength (e.g., \cite{madore2012}). In addition, there are currently no or insufficient infrared time-series data of bright (bright in the infrared) variable stars. We have been monitoring the same area as that of the photometric survey since October 2016 with a typical cadence of twice a month. We use 15~dithering mode for the variable star survey with the effective exposure time of 105~s. This monitoring survey will provide systematic time series data of the bright stars for the first time. We plan to continue this survey as long as possible.

\begin{figure*}
\centering
\includegraphics[scale=0.50,angle=0]{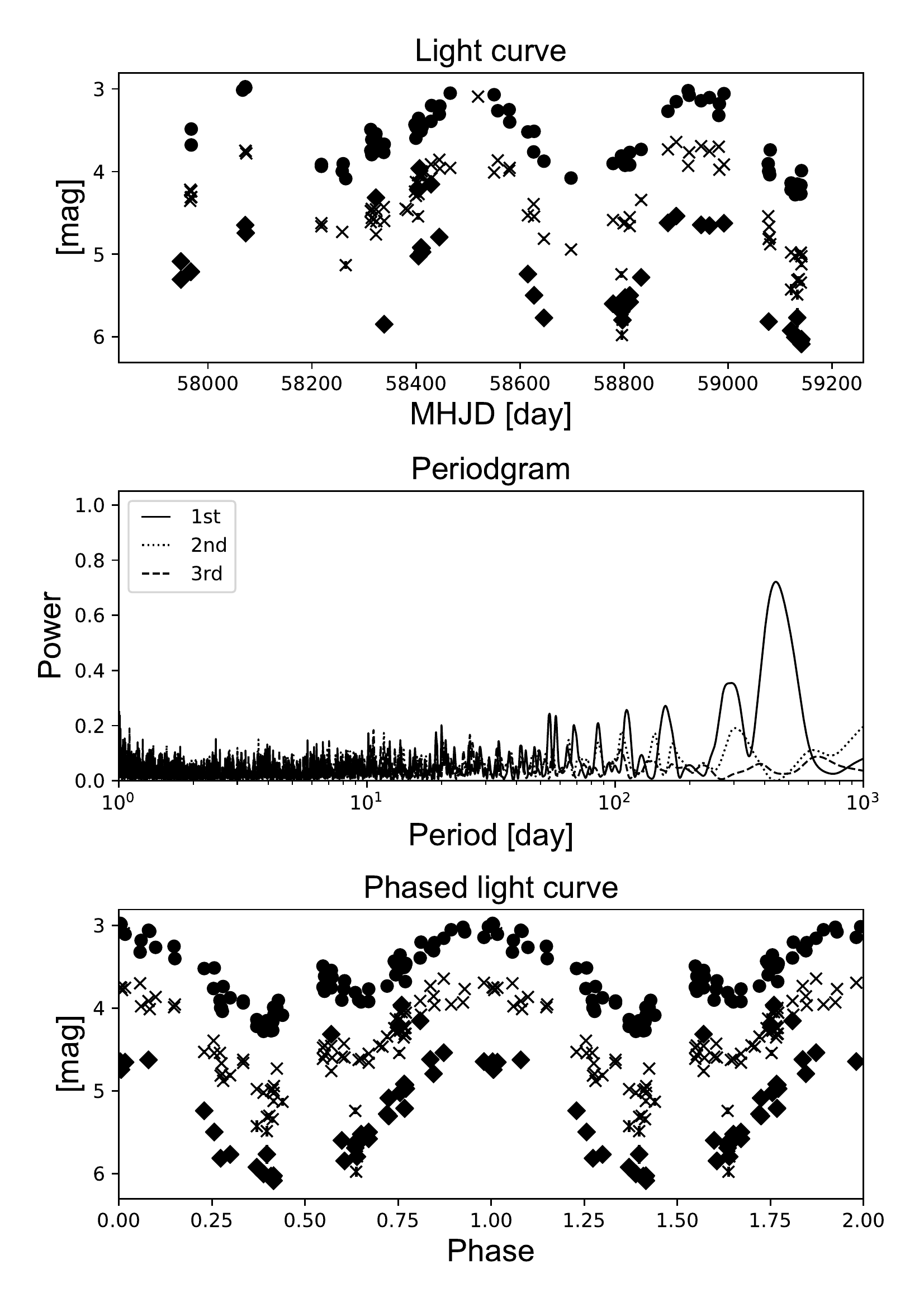} 
\includegraphics[scale=0.50,angle=0]{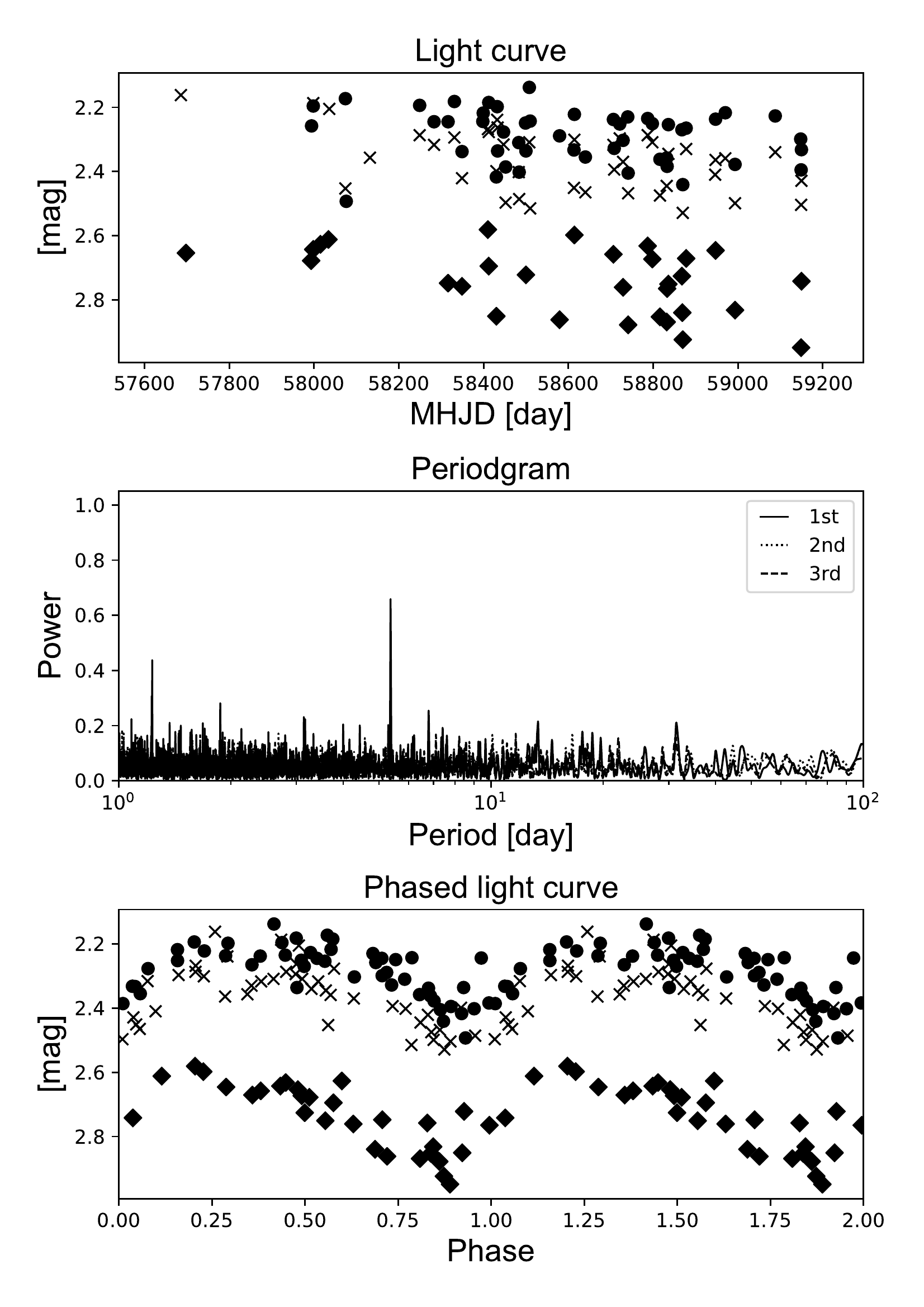} 
\caption{Light curves, Lomb-Scargle periodograms, and phased light curves of KK Sge (left) and $\delta$ Cep (right). Diamonds, crosses, and filled circles show $J$, $H$, and $K_{\rm s}$ bands, respectively. For most data points, error bars are smaller than the size of the symbols.}
\label{lightcurve}
\end{figure*}

To demonstrate our multi-epoch survey data, we show in Figure~\ref{lightcurve} two examples of the light curves, Lomb-Scargle periodograms (\cite{lomb1976}, \cite{scargle1982}), and phased (phased with the primary period) light curves for a Mira-like variable (KK Sge) and a classical Cepheid variable ($\delta$ Cep).

In both photometric and variable star surveys, we intentionally take slightly defocused images so that the full width at half maximum (FWHM) becomes 3 to 4 pixels in order to avoid under-sampling.

\section{Photometric calibration}
This section describes how we calibrate the photometry. Source detection and aperture photometry are performed using SExtractor (\cite{bertin1996}) to the combined images. In this section, $\lambda$ represents one of the photometric bands, $J, H$ or $K_{\rm s}$, and ${}^{\lambda}M_{\rm T}$ and ${}^{\lambda}m_{\rm T}$ refer to the calibrated and instrumental magnitudes in the TMMT system, respectively. Also, ${}^{\lambda}M_{\rm 2}$ refers to the 2MASS catalog magnitude for the photometric band $\lambda$.

We start by assuming that the following simple relationship exists between ${}^{\lambda}M_{\rm T}$ and ${}^{\lambda}m_{\rm T}$. 
\begin{equation}
\label{ca1}
{}^{\lambda}m_{\rm T} = {}^{\lambda}M_{\rm T} + {}^{\lambda}\gamma
\end{equation}
The calibration constant ${}^{\lambda}\gamma$ is determined as follows. 

\begin{figure}
 \begin{center}
  \includegraphics[angle=0,scale=0.5]{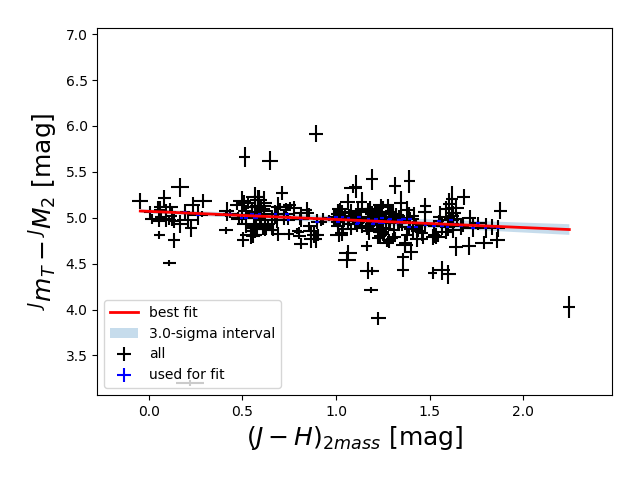} 
  \includegraphics[angle=0,scale=0.5]{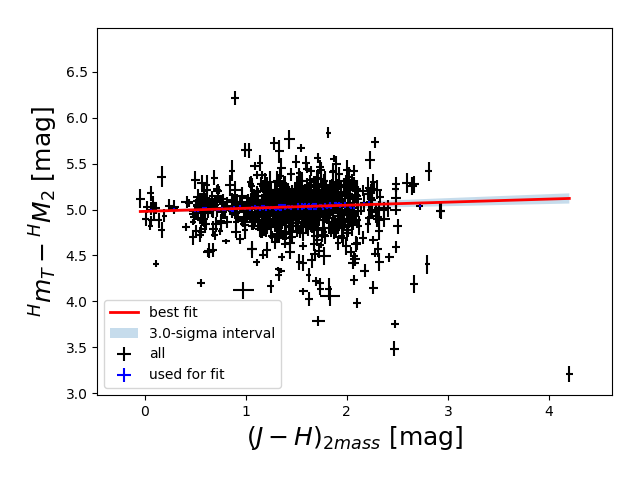} 
  \includegraphics[angle=0,scale=0.5]{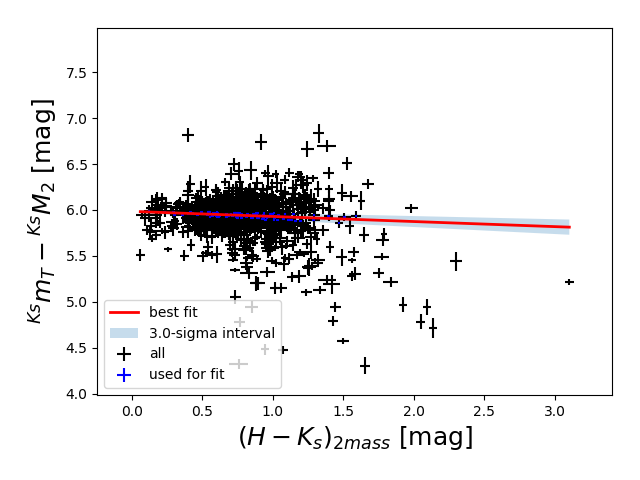} 
 \end{center}
\caption{An example of photometric calibration for one pointing data set. The black crosses indicate all 2MASS-TMMT pairs found in a FOV. The blue crosses show the pairs used for the regression. The best fit linear relation is shown by the solid red line. The cyan hatched area shows the 3~$\sigma$ interval of the best fit.}
\label{fig:calibration}
\end{figure}

The system transformation equations from the 2MASS system to the TMMT system can be defined in a form of 
\begin{equation}
\label{ca2}
{}^{\lambda}M_{\rm T} = {}^{\lambda}M_{\rm 2} + {}^{\lambda}a_{\beta} \times \beta + {}^{\lambda}b_{\beta},
\end{equation}
where $\beta$ is either of $(J-H)_{\textrm{2mass}}, (H-K_{\rm s})_{\textrm{2mass}}$ or $(J-K_{\rm s})_{\textrm{2mass}}$, and ${}^{\lambda}a_{\beta}$ and ${}^{\lambda}b_{\beta}$ are a slope and an intercept of the relation, respectively. Obviously, higher order terms arising from varying observing conditions are ignored here, but their effects would be small in the NIR. Combining equations (\ref{ca1}) and (\ref{ca2}), we can obtain the following equation.
\begin{equation}
\label{ca3}
{}^{\lambda}m_{\rm T} - {}^{\lambda}M_{\rm 2} = {}^{\lambda}a_{\beta} \times \beta + {}^{\lambda}b_{\beta} + {}^{\lambda}\gamma
\end{equation}
The slope ${}^{\lambda}a_{\beta}$ and intercept ${}^{\lambda}b_{\beta} + {}^{\lambda}\gamma$ can be determined by a linear regression to the difference of the TMMT instrumental magnitudes and the 2MASS catalog magnitudes as a function of the 2MASS color (see Figure~\ref{fig:calibration}). 

The intercept ${}^{\lambda}b_{\beta} + {}^{\lambda}\gamma$ cannot be resolved into each component. Since the ${}^{\lambda}b_{\beta}$ term in equations (\ref{ca2}) and (\ref{ca3}) is arbitrary, we choose ${}^{\lambda}b_{\beta} = 0$, so that the TMMT system is defined to have a same magnitude as 2MASS magnitude for a star with 2MASS color of zero. 
 
Thus, we can compute the calibration constant ${}^{\lambda}\gamma$ and obtain the calibrated TMMT system magnitudes. These equations are self-consistent and are good enough as long as discussions are completed within the TMMT system. Since the 2MASS system has become one of the most widely used NIR photometric systems today, it may be useful to derive the system and color conversion equations from the TMMT system to the 2MASS system. The transformation equations are beyond the scope of this paper and will be discussed in elsewhere.

The practical calibration procedure for each FOV and for each filter is as follows. First, we search for 2MASS stars within 10$^{''}$ from the position of a TMMT star. The search radius of 10$^{''}$ is chosen to be comparable to the size of the diffraction limited point source size in the $K_{\rm s}$-band (see Table~\ref{tab:spec}). Let us remind that we intentionally take slightly defocused images to avoid under-sampling. If there exists a 2MASS counterpart and its photometric error is smaller than 0.05~mag, we use the pair for the photometric calibration. If multiple 2MASS counterparts exist, the closest one is adopted, and others are discarded. Figure~\ref{fig:calibration} shows an example of the calibration process for one pointing dataset. After removing outliers, which are likely to be caused by mispairing, by iterating 3~$\sigma$ clipping, a linear relation is fitted to the remaining pairs by least squares method to obtain ${}^{\lambda}a_{\beta}$ and ${}^{\lambda}\gamma$ in equations (\ref{ca2}) and (\ref{ca3}). In this figure, it can be seen that there is a linear correlation between the 2MASS and the TMMT systems, as assumed in Equation~\ref{ca2}.

We admit that our calibration can be affected by large amplitude variable stars. The light variation amplitudes in the NIR wavelengths are generally small for most types of variable stars. Possible exceptions are Miras and OH/IR variables, whose peak to valley light variation in the $K_{\rm s}$-band can be as large as $\sim$ 2~mag (e.g., \cite{wood1998}). They are evolved red giants, which are intrinsically very bright in the NIR. In fact, many Miras and OH/IR variables have been detected in the TMMT survey. Nevertheless, such large amplitude variables should be outnumbered by non-variable red giants that are also bright in the NIR. Furthermore, as the large amplitude variables are likely to be removed in the 3~$\sigma$ clipping iteration processes, their contribution would be small.

\section{Astronomical performance}
\subsection{Astrometric performance}
\subsubsection{Astrometry}
The celestial coordinates of the sources detected by TMMT are calculated by referencing their positions to the 2MASS point source catalog\footnote{Note that the absolute astrometric precision of the 2MASS point source catalog is $\sim$0.07$^{\prime\prime}$ (\cite{skrutskie2006}).} by using SCAMP (\cite{bertin2006}). The typical two-dimensional root mean square value of the positional difference between the 2MASS and the TMMT coordinates is 4$^{\prime\prime}$ to 5$^{\prime\prime}$ (corresponding to 0.5 to 0.6 pixels) for all wavebands. As it is $\sim$ 1/10 FWHM, the astrometric accuracy is considered to be reasonable. As the equatorial coordinates in the 2MASS point source catalog are given in reference to the International Celestial Reference System (ICRS), our coordinates are thus based on the ICRS.

\begin{figure}
\begin{center}
\includegraphics[scale=0.43,angle=0]{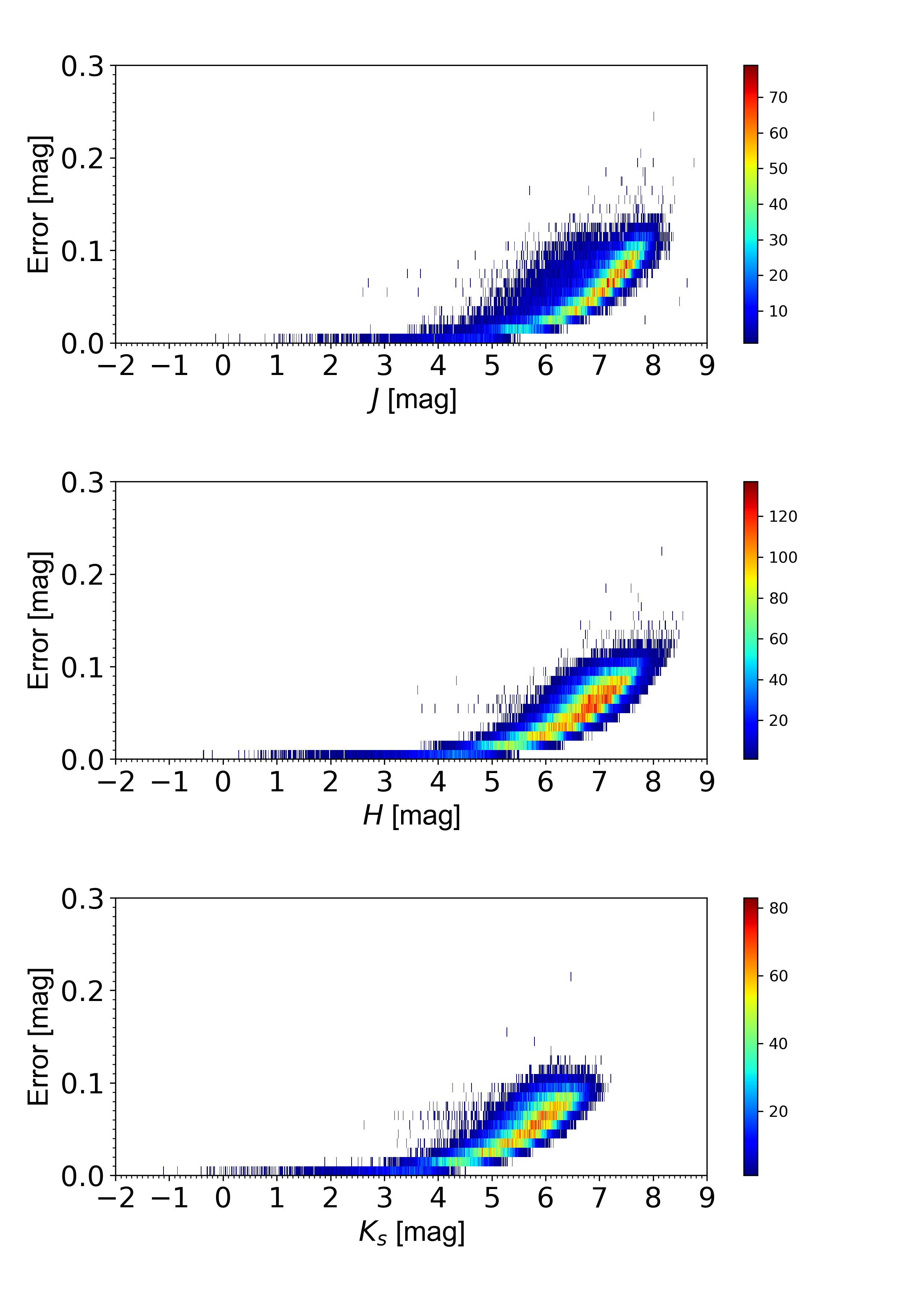} 
\end{center}
\caption{Heatmaps of the TMMT apparent magnitudes and their associated errors measured with SExtractor for the $J$, $H$, and $K_{\rm s}$ bands. The width of the bin is chosen to be 0.01~mag for both x and y axes. The color bar shows the number of sources in the 0.01 $\times$ 0.01~mag bin.}
\label{errorplot_tmmt}
\end{figure}

\subsubsection{Field of view}
The FOV of TMMT depends on the physical size of the detector and the focal length, which are summarized in Table~\ref{tab:spec}. Thanks to the large format of the detector, TMMT has a FOV as wide as 4.98$^{\circ}$ $\times$ 4.98$^{\circ}$ in the $K_{\rm s}$-band. The on-sky pixel scale and the array field of view obtained from the above astrometry are given in Table~\ref{tab:spec} for each band.

\subsection{Photometric performance}
\begin{table}
  \tbl{Magnitude zero point, zero-point flux density, and total throughput for each filter band.}{
  \begin{tabular}{lrrr}
      \hline
      Filter & $m_z$ & $F_0$ & $\tau$ \\ 
             & [mag] & [photons~$\mu$m$^{-1}$~s$^{-1}$~m$^{-2}$] & [$e^-$~photons$^{-1}$] \\ 
      \hline
      $J$ & 14.97 & 1.85 $\times$ $10^{10}$ & 0.64 \\
      $H$ & 14.63 & 8.38 $\times$ $10^{9}$ & 0.61 \\
      $K_{\rm s}$ & 13.98 & 4.74 $\times$ $10^{9}$ & 0.53 \\
      \hline
  \end{tabular}}\label{tab:throughput}
%\begin{tabnote}
%This is table note.
%\end{tabnote}
\end{table}

\subsubsection{Total throughput}
Here we define the total throughput $\tau$ in $e^-$~photon$^{-1}$ as the ratio of photoelectrons generated by the detector to the number of photons entering the aperture of the telescope. Then the $\tau$ can be written as
\begin{equation}
\label{eq:throughput}
\tau = \frac{g}{F_0 10^{-0.4m_z} \Delta \lambda A},
\end{equation}
where $g$ is the conversion gain in $e^-$~ADU$^{-1}$, $F_0$ the zero-point flux density in photons~s$^{-1}$~$\mu$m$^{-1}$~m$^{-2}$, $m_z$ the zero-point magnitude that corresponds to a photocurrent of 1~ADU~s$^{-1}$, $\Delta \lambda = \lambda_{\rm cut-off} - \lambda_{\rm cut-on}$ the bandwidth in $\mu$m and $A$ the collecting aperture in m$^2$. We adopt the synthetic Vega spectrum (\cite{cohen1992}) as the definition of zero-point flux density. Here, $\tau$ is a practical efficiency which includes atmospheric attenuation. It would be difficult to estimate separately the contribution of atmospheric attenuation, especially in the NIR, due to Forbes effect (\cite{manduca1979}, \cite{young1994}). 

The photometric calibration is performed for each field of view because $m_z$ in equation~\ref{eq:throughput} would change from time to time. Therefore, we demonstrate $\tau$ for the data of one pointing taken on January 26, 2019 (this was not chosen for any particular reason, but was chosen at random). The results and related quantities used in the calculations are summarized in Table~\ref{tab:throughput}. Throughput is a measure of the efficiency of an instrument. Thanks to the small number of optical elements of TMMT, the high efficiency is achieved.

%https://pdfs.semanticscholar.org/883e/d3ae4cf17fe410a0632747a1786b873da7da.pdf
\subsubsection{Typical photometric error}
Figure~\ref{errorplot_tmmt} shows the photometric errors given by SExtractor as a function of the apparent magnitude in the TMMT system for the whole stars detected in the northern sky photometric survey (see section~\ref{sec:survey}). Note that the errors indicate a lower limit of the true uncertainty, as it only takes into account photon and detector noises (we call them as random error, hereafter). The upward increase of the random error for fainter stars is clear, although there are some outliers that would be due to blending. In order to discuss the final photometric errors, it is necessary to evaluate the systematic errors arising from our calibration. As described in the previous section, we perform photometric calibration for each field of view. Therefore, we have to take into account systematic errors in the calibration process in addition to the random error in Figure~\ref{errorplot_tmmt}. The systematic error would vary from FOV to FOV because the observing conditions and the number of stars used in the calibration are different for each FOV. Usually, the calibration error is negligible for fainter stars, but it becomes comparable for bright stars with small random errors.

\begin{figure}
 \begin{center}
  \includegraphics[angle=0,scale=0.41]{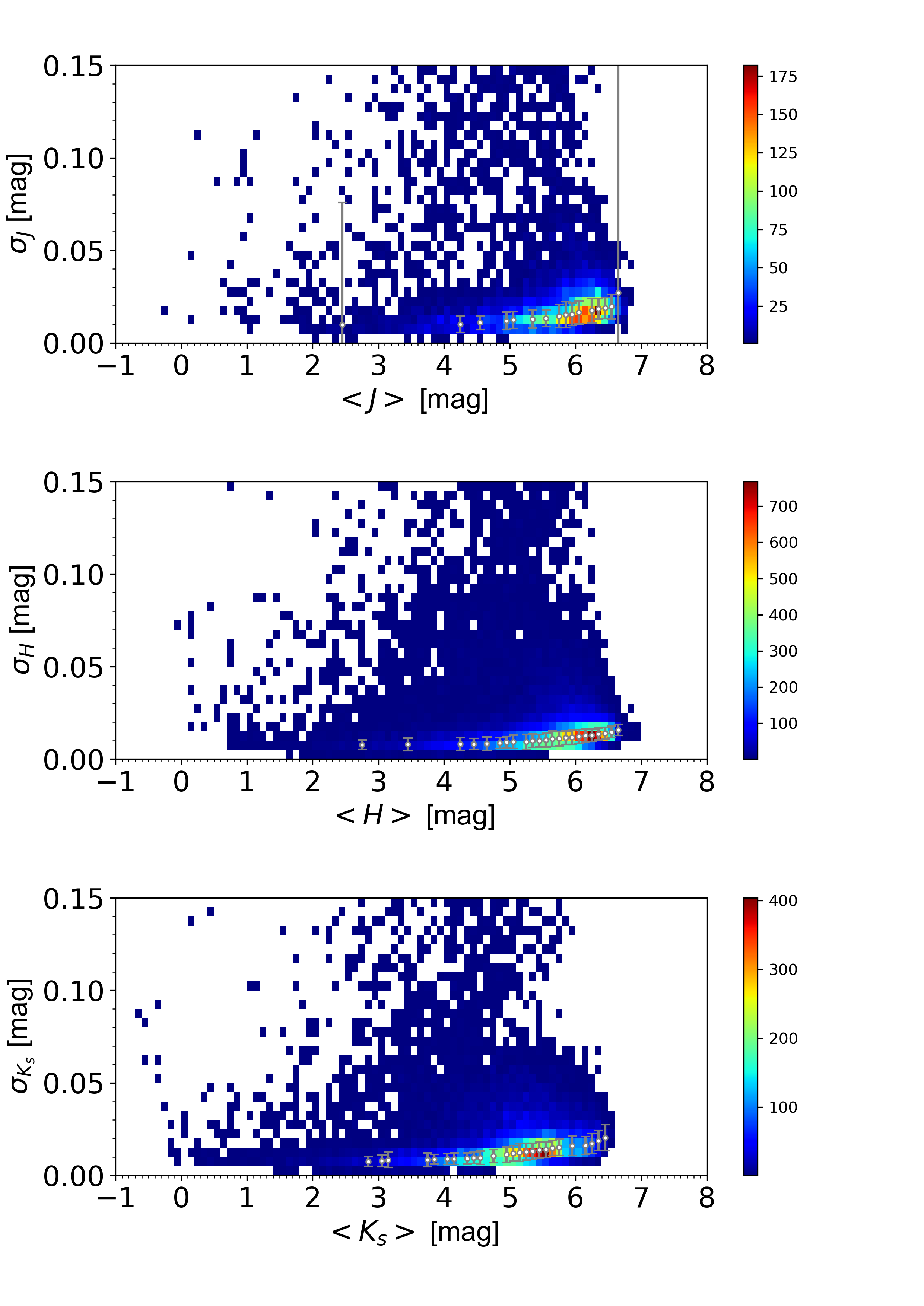} 
 \end{center}
\caption{Heatmaps of the uncertainty-weighted averages and their standard deviations of the calibrated time-series photometric data for all stars but known large amplitude variables (see the main text for details). The averages and standard deviations are indicated by gray circles with error bars. The width of the bin is chosen to be 0.1 and 0.005~mag for both $x$ and $y$ axes, respectively. The color bar shows the number of sources in the bin.}
\label{fig:precision}
\end{figure}

\subsubsection{Photometric precision}
The photometric precision can be assessed from the repeatability of observations. We use the time-series data obtained in the variable star survey (see section~\ref{sec:survey}) to calculate the uncertainty-weighted mean magnitude and its standard deviation for all detected stars. Here we denote the uncertainty-weighted mean magnitude in a waveband $\lambda$ as $\langle \lambda \rangle$ [mag] and its standard deviation as $\sigma_\lambda$ [mag]. 

Since the start of our variable star survey in 2016, we have been accumulating time-series data. A part of the survey data (for 60 fields of view) is used to calculate $\langle \lambda \rangle$ and $\sigma_\lambda$. The typical observational cadence is twice a month, and there are at least 30 independent measurements for each star.

The $\sigma_\lambda$ can be affected by many factors, such as photometric error, observing conditions that change moment-by-moment, and intrinsic photometric variability of stars. We identify known large amplitude variables, namely Miras, by crossmatching the positions of our samples with the VSX catalog (\cite{watson2006}), within a radius of 5$^{''}$. Then we remove the variable stars from the analysis.

The relations of $\langle \lambda \rangle$ and $\sigma_\lambda$ are shown in Figure~\ref{fig:precision}, which suggests a photometric precision of $\sim$1~\% for stars brighter than 5.0~mag. The high precision demonstrates that the photometric performance is very stable and that TMMT is capable of detecting bright variable stars with relatively small amplitudes.

\section{Summary}
It is very difficult to observe bright stars with modern large telescopes because of saturation in the detectors. The accurate NIR data of bright stars are absent or insufficient at present. To overcome this situation, we have developed a small telescope based on a 2048 $\times$ 2048 HgCdTe array. The telescope is called the Thirty MilliMeter Telescope (TMMT) after its effective aperture diameter. It has only two optical components (objective lens and dewar window) besides the Mauna Kea Observatories Near-Infrared $J, H$, and $K_{\rm s}$ filters. TMMT has been in use since October 2016 and we have carried out a photometric survey of the belt-like area along the northern Galactic plane. The same area has been monitored also. TMMT can provide photometry with an uncertainty of less than 5\% for stars brighter than 7, 6.5, and 6~mag in the $J$, $H$, and $K_{\rm s}$ bands, respectively. The repeatability of the photometric measurements is better than 1\% for bright stars. The high performances of TMMT meet our scientific goals to provide a NIR-bright point source catalog in the northern sky and their time-series data. The point source catalog will supplement the TMSS catalog for bright stars (Ita et al., in preparation).

\begin{ack}
The TMMT project is a collaboration between Tohoku University and the National Astronomical Observatory of Japan supported by the Grant-in-Aid for Scientific Research (C) No.18K03690 and (A) No.16H02158, and for Challenging Exploratory Research No.17K18789 from the Ministry of Education, Culture, Sports, Science and Technology (MEXT) of Japan. This work was also supported by the Sasakawa Scientific Research Grant No.27-202 from The Japan Science Society. This publication makes use of data products from the Two Micron All Sky Survey, which is a joint project of the University of Massachusetts and the Infrared Processing and Analysis Center/California Institute of Technology, funded by the National Aeronautics and Space Administration and the National Science Foundation. 
\end{ack}

%\appendix 
%\section*{Case of single paragraph}
%
%\section{Case of two or paragraphs}
%
%\section{Case of two or paragraphs}

%%%
% See the manual for the detail.
%%%

\end{document}